# A Search for Magnetized Quark Nuggets (MQNs), a Candidate for Dark Matter, Accumulating in Iron Ore

**J. Pace VanDevender [1,*], T. Sloan [2] and Michael Glissman [3]**

[1] MQN Collaboration, 10501 Lagrima De Oro Rd. NE, Albuquerque, NM 87111, USA
[2] Department of Physics, Lancaster University, Lancaster LA1 4YB, UK; t.sloan@lancaster.ac.uk
[3] Polymet Mining Inc., 6500 County Road 666, P.O. Box 475, Hoyt Lakes, MN 55750, USA; mike.glissman@newrangecoppernickel.com
* Correspondence: pace@vandevender.com; Tel.: +1-505-228-9998

**Abstract:** A search has been carried out for Magnetized Quark Nuggets (MQNs) accumulating in iron ore over geologic time. MQNs, which are theoretically consistent with the Standard Models of Physics and of Cosmology, have been suggested as dark-matter candidates. Indirect evidence of MQNs has been previously inferred from observations of magnetars and of non-meteorite impact craters. It is shown in this paper that MQNs can accumulate in taconite (iron ore) and be transferred into ferromagnetic rod-mill liners during processing of the ore. When the liners are recycled to make fresh steel, they are heated to higher than the Curie temperature so that their ferromagnetic properties are destroyed. The MQNs would then be released and fall into the ferromagnetic furnace bottom where they would be trapped. Three such furnace bottoms have been magnetically scanned to search for the magnetic anomalies consistent with trapped MQNs. The observed magnetic anomalies are equivalent to an accumulation rate of ~1 kg of MQNs per $1.2 \times 10^8$ kg of taconite ore processed. The results are consistent with MQNs but there could be other, unknown explanations. We propose an experiment and calculations to definitively test the MQN hypothesis for dark matter.

**Keywords:** cosmology; dark matter; quark nugget; magnetized quark nugget; MQN; MACRO; strange matter; strange quark matter

## 1. Introduction

On theoretical grounds, Bodmer [1], Witten [2] and Fahri and Jaffe [3] found that particles made of up, down, and strange quarks could meet the requirements for dark matter [4], the mysterious material holding galaxies together. Such a collection is called a quark nugget and should be more stable than normal matter since the three flavors of quarks provide three Fermi surfaces to fill, while normal matter composed of only up and down quarks have only two Fermi surfaces. Using a relativistic one-gluon exchange model, Tatsumi [5] found that quark nuggets could be ferromagnetic. He also showed that the exceedingly large magnetic field of magnetar pulsars is consistent with their having a ferromagnetic core of strange quark matter, supporting the hypothesis that ferromagnetic quark nuggets might exist.

Modeling the aggregation of ferromagnetic MQNs from their formation at ~65 μs after the beginning of the Universe, VanDevender et al. [6] found that the aggregation time from magnetic attraction was much less than the particle decay time of 0.26 ns [7]. Therefore, aggregation should have produced a broad mass distribution of stable MQNs as the Universe expanded in accord with the Standard Model of Cosmology. The magnetic field of magnetars [5,8] and observations from non-meteorite impacts [9–12] helps narrow the range of the surface magnetic-field parameter $B_o$ to 1.65 +/− 0.35 Tera Tesla (TT), which is approximately that at the surface of a neutron or proton. The corresponding computed mass range covers ~$10^{-23}$ kg to ~$10^{6}$ kg.

The Standard Model of Physics describes the fundamental particles, including quarks, electrons, etc. Baryons are particles composed of quarks; if they exist, MQNs are technically baryons [2] and are included in the Standard Model of Physics. Without compelling evidence for stable baryons beyond protons and neutrons, "baryonic matter' is often used as a synonym for atomic or normal matter. Since many observations show that dark matter is inconsistent with atomic matter, one often reads that dark matter cannot be baryonic. Therefore, the reader may conclude MQNs are inconsistent with dark matter. However, Jacobs, Starkman, and Lynn [4] have shown that nuclear-density baryonic matter with no significant electrical charge and with single masses between 0.055 and $10^{14}$ kg meet the theoretical and observational constraints of Cold Dark Matter (CDM)—unless the material converts normal matter to dark matter. They do not consider the broad mass distribution of MQNs nor their strong magnetic field. The work in [6] shows that >99.9999% of MQN mass falls within the allowed region for baryonic dark matter [4] and builds on [4] to show these two additional characteristics of MQNs are compatible with the conclusion of Jacobs, Starkman, and Lynn [4]. Of course, if MQNs convert normal matter to dark matter, the results reported in this paper cannot be attributed to MQNs. Therefore, the theory and discussions of [4,6] provide the theoretical foundation for stable MQNs consistent with dark matter and readers interested in that question are encouraged to examine that paper. However, unless MQNs are demonstrated to exist and to exist in sufficient abundance to account for dark matter, the discussion is premature. This paper provides additional evidence consistent with MQNs in numbers sufficient to their being dark matter, but the evidence is not dispositive. The proposed experiment described near the end of the paper should conclusively validate or invalidate the MQN hypothesis for dark matter and experimentally resolve the question.

DeRujula and Glashow [13] estimated that the gravitational force on unmagnetized quark nuggets with mass > 0.3 ng and at rest in rock would be sufficient to pull quark nuggets through rock to the center of the Earth. However, our COMSOL [14] simulations described in Appendix A, Method and validation of COMSOL Simulations, show the magnetic force on MQNs in magnetite ($Fe_3O_4$) veins would be sufficient to overcome gravity for MQNs with mass ≥ approximately 30 mg. If MQNs exist, they may have accumulated in iron ore over geologic time.

In this paper an experiment is described to search for MQNs accumulated in iron ore over ~$2 \times 10^9$ y or 2 Ga. The weakly magnetic iron ore releases the MQNs when ore is crushed in a rod mill consisting of a horizontal, rotating, steel-lined barrel in which thick steel rods continually fall through circulating ore and crush the ore into powder. Any MQNs would initially be picked up by the ferromagnetic rods but would then be released by the violent impact with the barrel liner, so that they are transferred into the ferromagnetic liners. These liners are periodically sent to a furnace to be melted down to make fresh steel. On heating to melting the MQNs would be released at the Curie temperature, fall through insulating fire bricks, and be trapped in the ferromagnetic steel bottom of the furnace. Three furnace bottoms were scanned to look for evidence of anomalous magnetic activity resulting from the presence of MQNs. Simulations are described which support this sequence of events.

Computer simulations, also described in Appendix A, show any accumulated MQNs with mass greater than about 0.1 kg would give a significant $|gradB_z|$, defined as the gradient in the absolute value of the vertical component of the magnetic field, on the bottom of these furnaces. We report the first measurements of the gradient in the magnetic field on the bottom of three such furnaces and compare the measurements with those of a control and with those of COMSOL computer simulations of MQNs in the furnace bottoms. The results are consistent with MQNs. However, the results are not dispositive since we cannot rule out the possibility of other, currently undiscovered explanations of the data. An experiment is proposed to definitively test the MQN hypothesis for dark matter.

Nuggets of dark matter with mass of ~0.3 kg would have an exceedingly small flux into the Earth, so it is not surprising why experiments looking for dark matter would not have detected one. However, one might wonder why ~0.3 kg MQNs, if they exist in ore, could have escaped detection during the processing of the ore. We show in the discussion section that the observed magnetic anomalies are equivalent to an accumulation rate of one ~0.3 kg MQN per 40 million kg of taconite ore processed. After crushing ore in a rock crusher to a mean size of ~0.03 kg, the process transporting the ore to the rod mills would have to check about 1.3 billion pieces to find one that weighs an extra ~0.3 kg—without disrupting the flow and productivity of the mill. Therefore, it is not surprising that MQNs have not been detected in the ore even if they exist in the inferred abundance.

We also show that processing ore in rod mills should concentrate MQNs into ~750 kg rod-mill liners that are melted down in furnaces for recycling. Our measurements of $|gradB_z|$ on furnace bottoms indicate there is about one ~0.3 kg MQN in about 4 liners, i.e., in about 3000 kg of liner steel. The corresponding fractional change in mass density is about 0.01%. We found the variation in the chemical composition of rod-mill rods (a steel with composition similar to that of rod-mill liners) accounts for the observed standard deviation in density of 0.18%, which is about 18 times the variation from an MQN in a liner. Consequently, it is not surprising that ~0.3 kg MQNS, if they exist in magnetite ore and recycled rod-mill liners, have not been previously observed in ore and liners.

However, our measurements are consistent with MQNs having been even more concentrated in the furnace bottoms. Over the 41-year operation of two furnaces, the mass of the furnace bottoms would be expected to increase by about 8 tons. That mass plus the 4-ton load of steel in a furnace is within the rated capacity of the furnace. The hydraulic motor tilting a furnace supplies more than enough torque to pour the steel without even slowing down the process. Imperatives of production at the foundry precluded removing a sample of the furnace bottom for directly weighing it and precluded indirectly estimating the furnace mass by the electricity consumed during the pour. Consequently, it is not surprising that these large-mass MQNs have not been previously detected in the ore or in the recycled rod-mill liners.

## 2. Materials and Methods

Since MQNs are strongly magnetized, they form a magnetopause when passing through a plasma, similar to the Earth's magnetopause with the solar wind [15]. Therefore, MQNs have a velocity-dependent interaction radius that is equal to the radius of their magnetopause and are slowed by their passage through ionized matter. The slowing-down process has been previously published; it is briefly summarized and extended in Appendix B, Calculation of acceptance fractions. The final result is the magnetopause radius $R_{mp}$ used for calculating the momentum transfer cross section $\pi R_{mp}^2$.

For an MQN of mass $m_{mqn}$, velocity $v_{mqn}$ moving through material of density $\rho_p$ with a drag coefficient $K \approx 1$, permeability $\mu_o$ of free space, $k = 10^{-18}$ m$^3$/kg = parameter inversely proportional to the dark matter density when the temperature of the Universe was ~100 MeV, and simulation parameter $B_o$ proportional to the surface magnetic field at the equator of the MQN, the momentum transfer cross section is

$$\pi R_{mp}^2 = \left(\frac{9\pi k^2 B_o^2 m_{mqn}^2}{8\mu_0 \rho_p v_{mqn}^2}\right)^{\frac{1}{3}} \quad (1)$$

Observations [10,11] limited the $B_o$ parameter to $B_o = 1.65 +/- 0.35$ TT. Therefore, we can compute the stopping of MQNs, within the assumptions of the model explained in Appendix B, Calculation of acceptance fractions.

The 21% uncertainty in $B_o$ gives ~16% uncertainty in stopping power and range in rock for MQN masses of 0.1 kg to 1.0 kg.

Theory [6] and observations [9–12] indicate MQNs should have a broad mass distribution, between $10^{-24}$ kg and ~$10^6$ kg. The mass distribution has been computed only to the precision of one order of magnitude in mass. Therefore, we will work with "decadal mass", which is defined as MQN masses with the same value of the function

Integer((log10 (mass)). Each decadal interval is represented by the approximate logarithmic mean of the interval, i.e., √10 kg ≈ 3.2 kg, times the minimum mass in the interval. For example, all MQNs with mass between 0.1 kg and ~1.0 kg are included as 0.32 kg decadal mass. In order to make the text easier to read, the term "decadal" may be dropped from the text. Unless otherwise stated, a mass of 3.2 times any power of 10 refers to the MQN mass in that decadal interval. We report that the 0.32 kg decadal mass interval is the most important one for detecting MQNs from iron ore because lower-mass MQNs do not produce sufficient magnetic field to be readily detectable above background and higher mass MQNs are fewer and are not as efficiently stopped by the rock above magnetite ore.

As already stated in the Introduction, MQNs are electrically neutral, nuclear-density baryonic matter, with masses that are not excluded by observational and theoretical constraints on dark matter [4,6]. Assuming the interstellar MQN mass density for all MQN masses is the mass density of the local dark matter halo = $4 \times 10^{-22}$ kg/m$^3$ [16,17], the number of MQNs/m$^3$ in interstellar space is given in Table A1 in Appendix B, Calculation of acceptance fractions, for the lower, middle, and upper values of $B_o$ and for MQN decadal-mass from 0.003 kg to 300 kg.

The fraction of incident MQNs deposited in ore (i.e., the acceptance fraction) and the limits of detection discussed in subsequent sections show the decadal mass of 0.32 kg (i.e., 0.1 to 0.9999 kg) is of the most interest in this paper. The fraction of the total interstellar mass density of the local dark matter halo contained in this one decadal mass interval is $24 \times 10^{-6}$, $5 \times 10^{-6}$, and $1.2 \times 10^{-6}$ for MQN mass distributions calculated [6] for $B_o$ = 1.3 TT, 1.65 TT, and 2.0 TT, respectively.

The uncertainty in the MQN number density arising from the uncertainty in the value of $B_o$ is represented in the last column of Table A1 as the variation in the $\mathrm{Log}_{10}$ of the MQN number density. The upper and lower uncertainty factors are, respectively, ~4.8 larger and ~1/4th as large as the value at $B_o$ = 1.65 TT. This uncertainty is obviously much larger than the +/− 16% uncertainty in the stopping power.

We differentiate between interstellar MQNs and solar MQNs. If all dark matter is composed of MQNs, interstellar MQNs form the dark matter halo of the galaxy. The velocity distribution of interstellar MQNs is computed in [6] and is found to be very cold because the process of magnetic attraction and aggregation of ferromagnetic MQNs means that the velocity of an aggregated MQN is the center-of-mass velocity of the two colliding MQNs. Since the center of mass velocity is less than either of the MQN velocities, each aggregation successively cools the distribution. Since gravitational acceleration is independent of mass, the gravitational effects that give the final thermal velocity distribution for CDM should give the same velocity distribution for interstellar MQNs. Therefore, we assume interstellar MQNs near Earth have the velocity distribution found in computer simulations [16] of cold dark matter (CDM): approximately Maxwellian thermal velocity plus the streaming velocity into the solar system consistent with the solar system's motion about the galactic center.

Solar MQNs are different. They have been slowed below escape velocity from the solar system by passing through just enough of the Sun's plasma and then scattering off the combined gravitational field of the Sun and planets to avoid being captured by the Sun or ejected out of the solar system. Since gravitational acceleration is independent of mass, the MQNs near Earth must have orbits like near-Earth asteroids and so they must have a velocity distribution like the measured velocity distribution of near-Earth asteroids [18–20].

A Monte Carlo technique, described in detail in Reference [12] and applied to MQN stopping in ore in Appendix B, Calculation of acceptance fractions, is used to sample the phase space distribution for MQNs incident over $2\pi$ steradians solid angle, with the vector sum of the streaming velocity and randomly oriented "thermal" velocity described above, and slowed down in their path through the exponentially increasing density distribution of the Earth's atmosphere when coming from space and through taconite ore with density of 2700 kg/m$^3$. The depth at which each MQN in the sample stops is stored. The program for the Monte Carlo calculation is available in the data upload accompanying this paper.

Since surface mines for magnetite and taconite are rarely more than 200 m deep, the fraction stopped in 100 m increments between 0 and 200 m is sufficient for our purposes. We include 0 to 500 m for use in assessing other deposits of iron ore.

The results, shown in Table A2, show the acceptance fractions are large for 0.003 kg masses and fall by ~80% for each decade increase in MQN mass. Only ~3% of MQNs with 0.3 kg decadal mass are stopped in the first 200 m of rock. If the magnetite-bearing formation has been covered by 1 to 2 km of non-ferromagnetic rock for half of the time since its formation, the 3% fraction is increased to ~58% to 99.7%, respectively, so 3% is a conservative estimate but could be a factor of 20 to 30 too low.

Equation (2) combines the interstellar number densities $n_{IS}$ of interstellar MQNs of decadal mass $m_{dec}$ in Table A1 and the acceptance fractions $f_{acc}$ in Table A2 to calculate the rate $\frac{dF_{mqn}}{dt}$ of accumulating interstellar MQNs of decadal mass $m_{dec}$ per gigaton (Gt =$10^{12}$ kg) of taconite ore per gigayear (Ga = $10^9$ y) of accumulation time as a function of decadal MQN mass for $B_o$ = 1.65 TT. The units are chosen to represent convenient industrial-scale quantities of mined ore and typical geologic accumulation times.

$$\frac{dF_{mqn}}{dt} = 0.7 n_{IS} m_{dec} v_{mqn} \frac{f_{acc}}{100} \frac{10^{12}}{2700} 3.154 \times 10^{16} = 5.8 \times 10^{22} n_{IS} m_{dec} v_{mqn} f_{acc} \frac{kg}{Ga}, \tag{2}$$

in which the factor of 0.7 is the fraction of the MQNs directed downwards towards the Earth for the velocity distribution given by the streaming plus thermal velocities, $f_{acc}$ is the is the fraction of the MQNs which stop in the rock per 100 m depth interval and the factor $\frac{f_{acc}}{100}$ gives the fraction of the MQNs which stop in the rock per meter depth, the factor $\frac{10^{12}}{2700}$ gives the number of cubic meters of rock with mass density 2700 kg/m$^3$ in a gigaton of rock, and the factor 3.154 × $10^{16}$ is the number of seconds in a Ga. Interstellar MQNs are assumed to have mean speed $v_{mqn}$ = 2.3 × $10^5$ m/s. The results are shown in Table A3.

For the case of $m_{dec}$ = 0.3 kg and MQNs from 0 to 200 m collecting in a 50 m thick taconite deposit at ~150 to 200 m depth, the accumulation rate is ~0.8 kg/Gt/Ga. Since taconite in the Iron Range of Minnesota USA may have been accumulating MQNs for 1.8 Ga, a gigaton of ore would be expected to yield only ~1.4 kg of ~0.3 kg MQNs with upper and lower uncertainty factors (in the right most column of Table A1), respectively, ~5 and 1/4th from the uncertainty in $B_o$. Even with the processing of taconite consolidating these MQNs, as discussed in subsequent sections, this accumulation rate would make detecting interstellar MQNs very challenging.

In principle, MQNs can accumulate in the solar system and provide an additional flux at lower velocity and shorter range in rock. We call these "solar MQNs" to differentiate them from interstellar MQNs. Xu and Siegel [21] computed the scattering of dark matter by individual planets into solar orbits. They estimate the accumulated mass density of dark matter in the solar system is ~$10^{4+/-1}$ times the background mass density of the dark matter halo if the dark matter were accumulated without losses for 4.5 Ga, the age of our solar system.

The calculation by Xu and Siegel involves only gravitational forces, so it is appropriate for all dark matter. Unlike most dark-matter candidates, MQNs also interact through their magnetopause, as discussed above, to slow down during passage through normal matter. Therefore, some interstellar MQNs would pass through just enough of the Sun's corona and chromosphere at grazing incidence to slow their velocity to below the escape velocity from the solar system and some of those would scatter from the combined gravitational field of the Sun and planets to be trapped in long-lifetime orbits. Therefore, we include the possibility of solar MQNs in the analysis with a Maxwell-Boltzmann velocity distribution peaking at 30 km/s, consistent with the velocity distribution of meteorites impacting Earth [20]. We assume the mass distributions of Table A1 apply to solar MQNs and compute their acceptance fractions with the same Monte Carlo technique applied to interstellar MQNs. The results are shown in Table A4.

Acceptance fractions for solar MQNs in Table A4 are significantly larger than they are for interstellar MQNs in Table A2. For the case of most interest, 0.3 kg decadal mass and 0 to 200 m depth, the acceptance is 78%, which is ~25 times larger than for interstellar MQNs. If the magnetite-bearing formation has been covered by 1 to 2 km of non-ferromagnetic rock for half of the time since its formation, the 78% fraction is increased to ~100%, respectively, so 78% is a conservative estimate and the uncertainty in the geological history is low.

As was performed above for interstellar MQNs in Table A3, the acceptance fractions for solar MQNs and the MQN mass distribution are combined with Equation (2) to estimate the accumulated number of solar MQNs per unit volume of taconite ore per Ga of accumulation time as a function of decadal MQN mass—if the number density of solar MQNs is the same as it is for interstellar MQNs in Table A1, i.e., we assume an enhancement factor of only 1 for this calculation. The results are shown in Table A5.

The larger acceptance factor for solar MQNs is largely offset by the lower velocity. For the case of $m_{dec}$ = 0.3 kg and MQNs from 0 to 200 m collecting in the 50 m thick taconite deposit at ~150 to 200 m depth, the accumulation rate is ~2.5 kg/Gt/Ga, only ~3 times the rate for interstellar MQNs if the solar MQN number density equals the interstellar number density. Potential enhancement of the solar MQN number density will be considered in the Discussion section.

The Mesabi Iron Range in Minnesota, USA, is rich in magnetite and taconite, which is a lower grade iron ore consisting of magnetite and non-magnetic quartzite. The formation is typically ~200 m thick and is composed of ~150 m thick non-magnetic Virginia formation on top of ~50 m thick taconite. Therefore, we consider the acceptance fraction of MQNs stopping in the first 200 m of rock and searched for evidence of MQNs from those taconite mines. Acceptance fractions for up to 500 m of rock above the iron ore are included in Appendix B to illustrate the sensitivity of the choice.

Collecting MQNs would require they remain in the magnetite ore as it is processed from the mine to the rod mills, where they can magnetically accumulate in steel hardware. Perhaps 10% of taconite ore has magnetite in cm-scale thick veins running for several centimeters to meters in length. The rest is in mm-scale deposits (granular iron formation) separated by non-magnetic minerals. Therefore, we calculated the depth of the potential well confining MQNs to magnetite grains of mm scale to cm scale. The calculation is explained in Appendix C, Holding force of steel and magnetite on embedded MQNs.

Since magnetite is a semi-conductor, its electrical conductivity of 2.2 × $10^4$ S/m [22] is too low for eddy-current braking to be significant and was not included. The results are presented as the height $h_{eject}$ of a free-fall drop in meters that will eject MQNs to well beyond the magnetic attraction of a sphere of magnetite, as a function of MQN mass $m$ in kg and magnetite radius $r_{magnetite}$ in meters:

$$h_{eject} = 6.0 + 0.354m^{-0.272} + 0.455ln\left(r_{magnetite}\right). \quad (3)$$

The results are weakly dependent on MQN mass and magnetite radius. For example, a drop of 3.3 to 3.0 m will eject 0.1 to 1.0 kg MQNs from 0.625 mm radius magnetite. A drop of 4.3 to 4.4 m will eject 0.1 to 1.0 kg MQNs from 5 mm radius magnetite. A mining process that limits drop heights to <3.0 m, or more conservatively <2 m, should retain any MQNs until they reach a rod mill, where the magnetite is crushed to a fine powder and any MQNs would be attracted into the high-permeability steel.

We measured the electrical conductivity of rods and liners to be 9.3 × $10^6$ S/m +/− 0.14 × $10^6$ S/m, which is ~ a factor of ~400 larger than the 2.2 × $10^4$ S/m electrical conductivity of magnetite [22]. The difference is sufficient to make eddy-current forces important for slowing MQN motion through rods. In addition, an MQN only has to be ejected out of the surface of the rod and into the steel of a liner to be transferred—not ejected to a large distance from the rod, as required for ejection from magnetite. Neglecting the additional complication of rod rotation, we find that a 0.32 kg MQN would be ejected from a rod by a drop height $h_{eject_rods}$ ~0.62 m without eddy-current forces and 0.68 m with eddy current forces. Our limited simulations provide an approximate mass dependence of

$$h_{eject_{rods}} = 0.54m^{-0.20}, \quad (4)$$

in which the exponent of −0.2 is similar to the exponent of −0.272 in Equation (3) for magnetite.

The six producing taconite mines in the Iron Range of Minnesota USA currently process about 0.12 gigatons of ore per year. Taconite ore is extracted with explosives and transported as boulders to rock crushers and then to mills for grinding into powder. For example, they have historically been crushed to ~2 cm diameter for processing in rod mills, where ~10 cm diameter steel rods grind the ore to powder.

If MQNs are transported through the rock crushers and into rod mills, our COMSOL computer simulations and laboratory emulations show that MQNs would be transferred to the higher permeability steel rods. On each revolution of a rod mill, some rod-mill rods are in the cascade zone (https://www.911metallurgist.com/blog/rod-mills, last accessed 6 January 2024) and are tossed at high-speed into the steel liners. Our COMSOL simulations of this process indicate the resulting impact is sufficient to transfer MQNs from rods to liners. Therefore, we conclude that any MQNs passing into rod mills would be accumulated in the steel liners.

Rod mills usually have a polymer or rubber backing separating the liner from the shell to absorb shock and to mitigate abrasion (by magnetite–water slurry) of bolts holding the liner to the shell. The backing also isolates the liner from the shell magnetically, so MQNs accumulated in the liner cannot reach the shell.

When rod-mill liners wear out, they are sent to a foundry where they are melted, combined with scrap metal and chemicals to produce the desired alloy for new liners and sold back to the mills. As the steel is heated, it transitions from ferromagnetic to paramagnetic state and MQNs would fall to the 2.54 cm thick steel bottom of the furnace, where they would accumulate for the multi-decade lifetime of a furnace. Therefore, these furnaces inadvertently serve as MQN collectors, concentrating any MQNs from several magnetite and taconite mines over decades into one location.

## 3. Results

### *3.1. Scanning Furnace Bottoms for MQN Signatures*

Many of the liners used in processing magnetite and taconite ore in the Mesabi Iron Range of Minnesota are melted in the three electric-arc furnaces of ME Elecmetal Inc. in Duluth, MN, USA. Management graciously allowed us access to the furnaces after operating hours to look for the characteristic magnetic signature of MQNs.

Computer simulations described in the next section showed that the distance between the bottom air–steel interface and an MQN's equilibrium position (from magnetic and gravitational forces) inside a 2.54 cm thick, A36-steel furnace bottom increases with increasing MQN mass. Therefore, the larger the MQN, the more effective its magnetic field is shielded by the steel. Detecting MQNs above the natural background of Earth's magnetic field and the steel's aggregated magnetic domains is difficult.

The industrial environment shown in Figure 1, in which the furnaces reside, presents special challenges. The furnace bottoms are convex surfaces and are at ~180 °C temperature. Scanning the furnace bottoms at a well defined pitch with a single magnetometer, as one might do in a laboratory environment, was not possible.

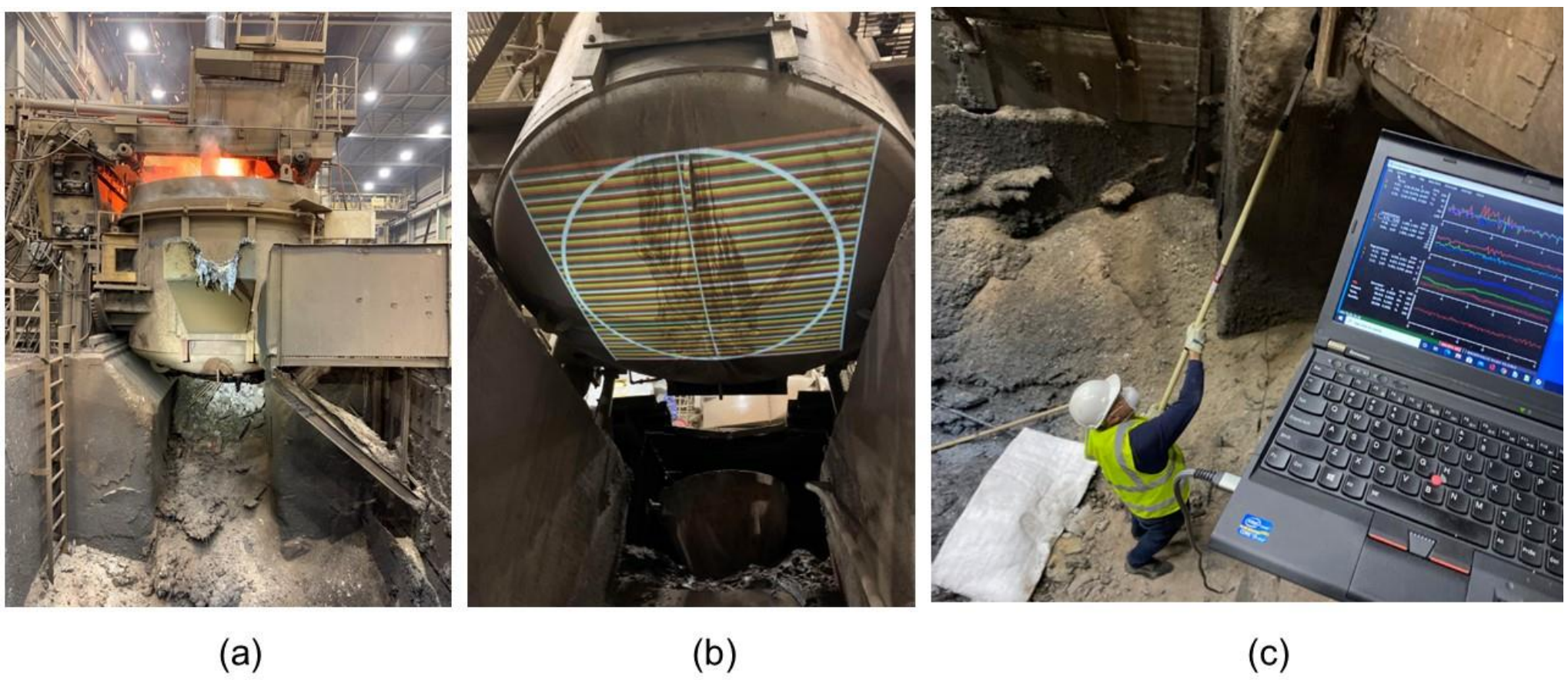

**Figure 1.** Photos illustrating the challenges and method of recording the magnetic field data on furnaces surfaces in the industrial environment. (**a**) View of ~3.8 m diameter furnace 1 in operation, (**b**) furnace 1 bottom with projected pattern to guide manual scanning, and (**c**) first author (JPV) in the pit scanning the B-field of the furnace shell in preparation for scanning the furnace bottom.

Computer simulations showed that scanning the bottom with two magnetometers separated by 1 cm and processing the data to give the distribution in the absolute value of the gradient in the vertical field component $B_z$,

$$|gradB_z| \equiv \frac{(B_2 - B_1)}{0.01} T/m \quad (5)$$

provides a practical method of distinguishing accumulated MQNs from the natural background.

A commercially available microsystem that included two 3-axis magnetometers $B_1$ and $B_2$ spaced 1 cm apart (ruggedized Micro Altitude Heading Reference System (uAHRS) by Inertial Sense Inc., Provo, UT, USA with their EvalTool software 1.8.4 b101_2021_03_30_172532) was used to scan the magnetic field profiles, as illustrated in Figure 2. The magnetic field readings of the B1 and B2 magnetometers were automatically recorded at slightly different times. The data of B2 was time synchronized to the times of B1 by interpolating the B2 data to give the B2 reading at the same time as the nearest B1 reading, thereby time synchronizing the two measurements of magnetic field separated by a constant 1 cm distance. Then, $|gradB_z|$ was calculated from Equation (5).

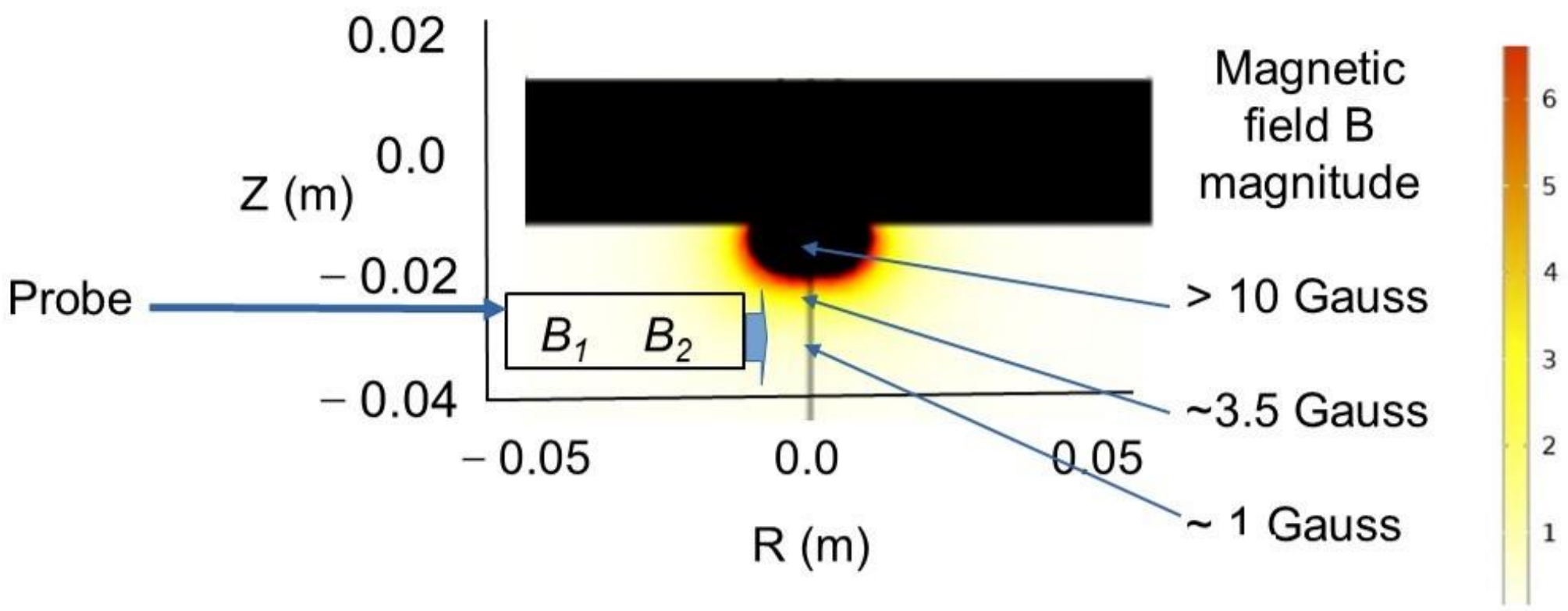

**Figure 2.** The probe with two magnetometers $B_1$ and $B_2$ is shown passing across the region of increased magnetic field near a simulated MQN inside a 2.54 cm thick steel plate.

The microsystem was mounted between two aluminum rollers on a spring-mounted platform that allowed the magnetometers at the center of the ruggedized package to be smoothly moved 7.5 +/− 1.0 mm from a surface and perpendicular to the vector connecting the two magnetometers. The microsystem was thermally isolated from the scanned surface by ~2 mm of silicone and a ~2 mm air gap.

The center 2.4 m diameter area of all three furnaces were scanned with the microsystem by projecting a colored pattern of parallel lines on the bottom of each furnace, as shown in Figure 1b, and moving the microsystem probe along each line at approximately constant speed and then manually advancing it to the next line at the end of each scan. This method sampled the entire 2.4 m diameter area even though the manual sampling rate was only approximately constant. Having the two probes at a fixed separation of 1 cm allowed $gradB_z$ data to be accurately recorded at approximately 35,000 imprecisely-defined locations (x,y) for each furnace. We interpreted the data as accurate measurements of $gradB_z$ with approximately random sampling within the scanned area.

Maps of scans were prepared by plotting y(m) + 28.5 ($T^{-1}$ $m^2$) × $gradB_z$ (T $m^{-1}$) versus x(m), in which the 28.5 ($T^{-1}$ $m^2$) multiplier was varied to adjust the apparent amplitude of $gradB_z$. Results for furnaces 1 and 2 are shown in Figure 3a. The plots indicated the approximate spatial distribution of $gradB_z$ even though the exact position of the probe for each measurement was not accurately determined. The raw data files and the program used to synchronize the $B_z$(t) measurements, compute $gradB_z$ and plot the data are available for examination in the data archive.

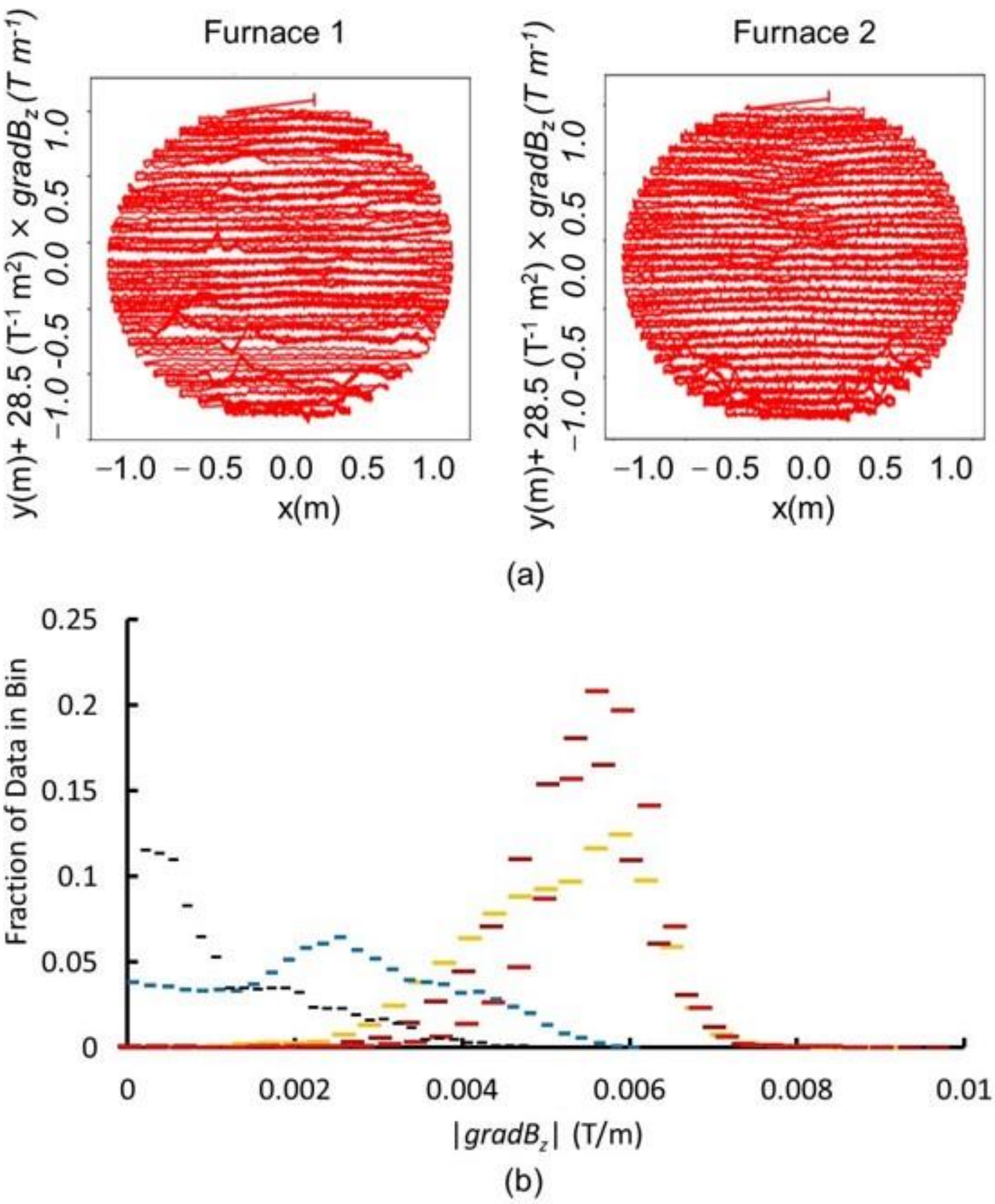

**Figure 3.** (**a**) Raster plots of y(m) + 28.5 ($T^{-1} m^2$) × *gradB*$_z$ (T $m^{-1}$) versus x(m) for furnace 1 (**left**) and furnace 2 (**right**) illustrate the approximate spatial distributions of *gradB*$_z$ and (**b**) histograms of the distribution of |*gradB*$_z$|, the absolute value of *gradB*$_z$, color coded as furnace 1 (orange), furnace 2 first scan (dark red), furnace 2 s scan (bright red), furnace 3 (blue), and control at 166 °C (black).

As illustrated in Figure 3b, histograms are difficult to follow when multiple data sets are plotted in the same graph for comparison, and the arbitrary width of the histogram bins reduces the resolution of the data. Therefore, we have elected to present the n data points ($x_i$,$y_i$) in its integral form, i.e., in a set that is ordered from the lowest (i = 1) to highest (i = n) value of |*gradB*$_z$|$_i$, so that $y_i$ = i/n, expressed as %, for $x_i$ = |*gradB*$_z$|$_i$. The corresponding plots are shown as "% of measurements < |*gradB*$_z$|" versus |*gradB*$_z$|. Therefore, for each data set, y ranges from 0% to 100% as x ranges from the minimum to the maximum value of |*gradB*$_z$| for that data set. The resulting plots are shown in Figure 4a.

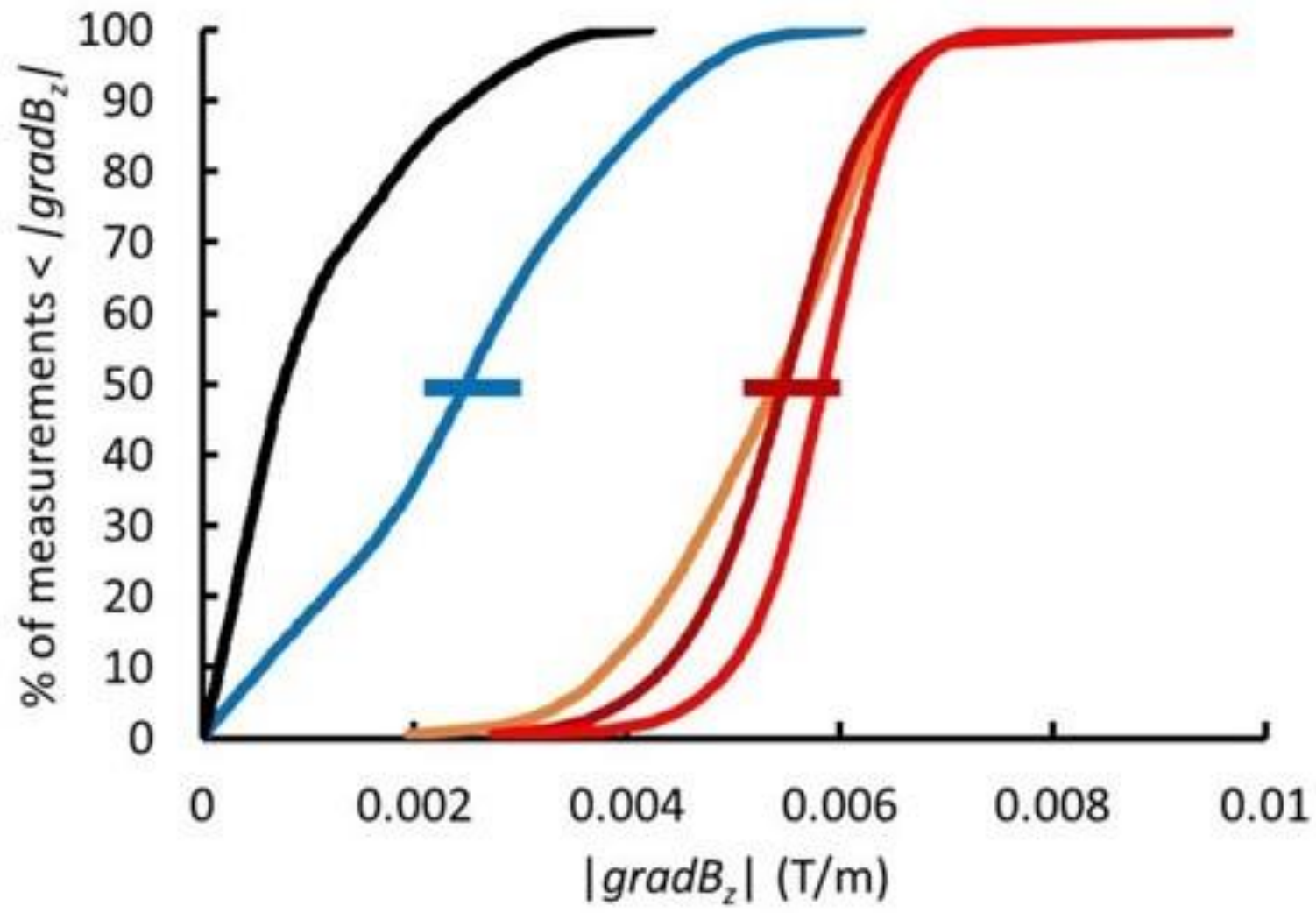

(a)

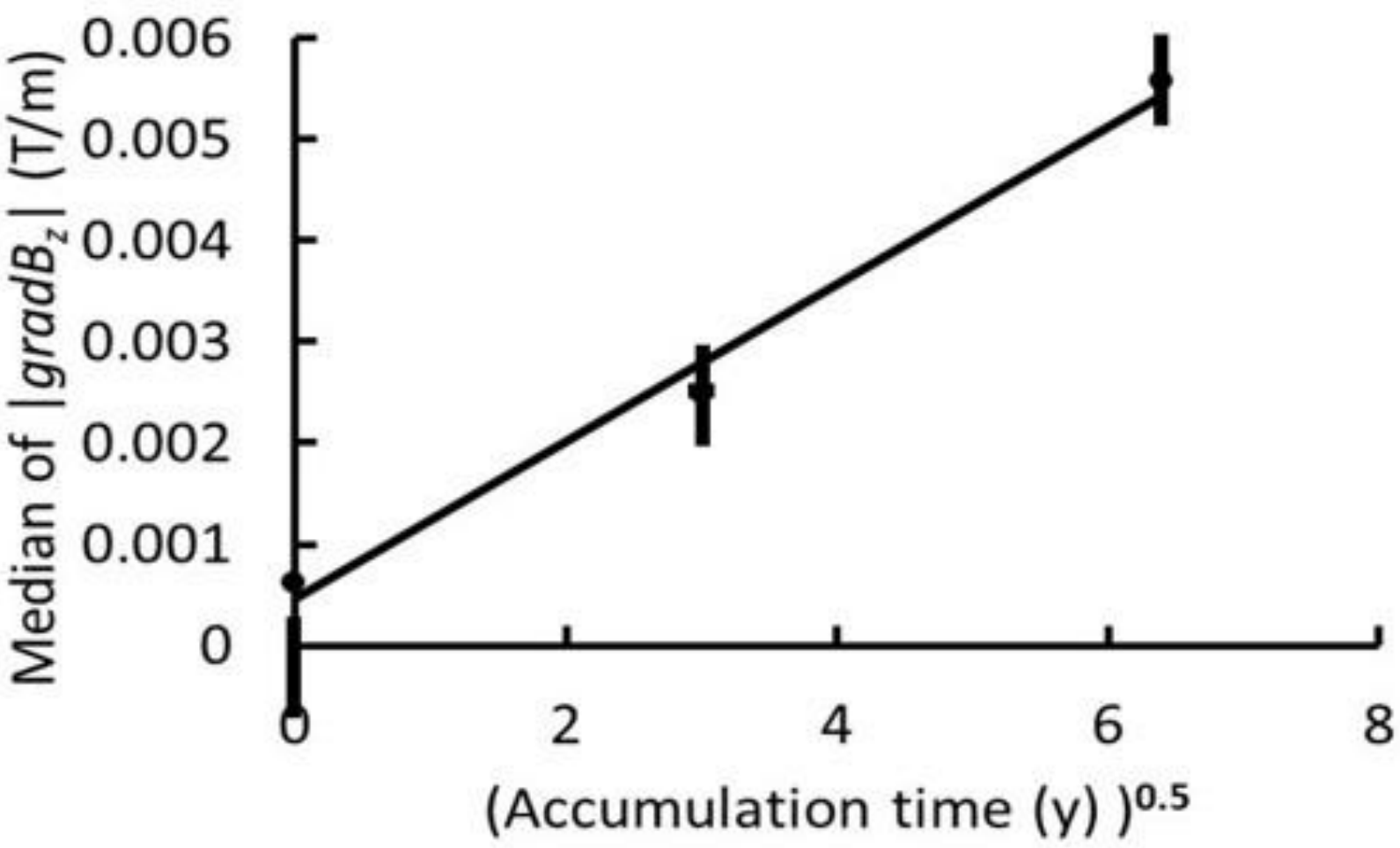

(b)

**Figure 4.** (**a**) Cumulative distributions of $|gradB_z|$ measurements on the three furnaces and the control: furnace 1 (orange), furnace 2 first scan (dark red), furnace 2 s scan (bright red), furnace 3 (blue), and control at 166 °C (black). (**b**) Plot of median of the $|gradB_z|$ distributions versus the square root of the age of the furnace; the linear relationship is consistent with a constant accumulation rate of whatever is causing the $|gradB_z|$ distributions. Error bars represent +/− 2$\sigma$ calculated uncertainty. The standard deviation $\sigma_x$ in the value of $|gradB_z|$ (i.e., ×coordinate) at the median (50%) for the three data sets from 41-year-old furnaces 1 and 2 in (**a**) was calculated and assumed to be representative of the corresponding uncertainty for the 9-year-old furnace 3 in (**a**). The standard deviation $\sigma_y \approx 0.3\%$ in the median position (i.e., y coordinate) in (**a**) was calculated from counting statistics for a binomial distribution with ~35,000 data points, which is typical for these curves, and for the distributions for furnaces 1 and 2 shown in Figure 3b. The error bar in the y coordinate is ~one third of the line width, so it does display. However, the paucity of runs and the unquantifiable uncertainties in the data make $\sigma$ a lower limit to the uncertainty as a measure of confidence level. We invite the reader to interpret them as only qualitatively indicative of the real uncertainty and as consistent with the difference between the median of the data from the 41-year-old furnaces 1 and 2 and the median of the data from the 9-year-old furnace 3. The error bars in (**b**) were derived from those in (**a**) and from the +/− 0.5 y uncertainty in the operational age of the respective furnaces.

Furnaces 1 and 3 were scanned once and furnace 2 was scanned twice to assess reproducibility. The results are shown in Figure 4a with the scan from the 166 °C scan of the control, described below, shown for comparison.

At the time of the scans on 4 December 2021, furnaces 1 and 2 had been in operation for 41 years, and furnace 3 had been in operation for 9 years, during which all furnaces shared the workload evenly. All three furnaces were manufactured by Whiting Furnaces and appear to be the same design and were made of the same A36 steel. The medians of the two scans of furnace 2 and the single scan of furnace 1, i.e., the two furnaces in operation for 41 years, agree within approximately +/− 4% (i.e., 0.000234 T/m).

The median for furnace 3 is 0.0025 T/m (operating for 9 years), and the median for furnaces 1 and 2 is 0.00557 T/m +/− 0.000234 T/m (each operating for 41 years). The median of the data for the control, presented below, is 0.00065 T/m. A constant accumulation rate over a fixed area means the number of MQNs per unit area increases linearly with time t. Since the number per unit area equals 1 MQN/$d^2$ for MQN separation d, 1/$d^2$ scales as t, and 1/d scales as $t^{0.5}$. Since $|gradB_z|$ = the change in $B_z$ per unit length, $|gradB_z|$ scales as $t^{0.5}$ for a constant accumulation rate, as it does in Figure 4b, consistent with the hypothesis that MQNs have accumulated in the mill liners from the processed taconite at a nearly constant rate.

An approximately 60-year-old, 12.7 mm thick, 0.86 m by 0.46 m, mild steel plate that was never exposed to rod-mill liners was used as a control. The temperature of the furnace bottoms was ~177 °C during operation and between 121 °C and 150 °C at the time of the scans. We scanned the center 0.76 m by 0.42 m area of the control plate as a function of temperature. Each scan of the control plate and furnace bottom recorded, respectively, 8500 +/− 500 and 33,800 +/− 1800 measurements. Plotting the cumulative distribution as % of measurements < $|gradB_z|$ versus $|gradB_z|$ provided a convenient way to visualize the data. The results of the scans of the control are shown in Figure 5.

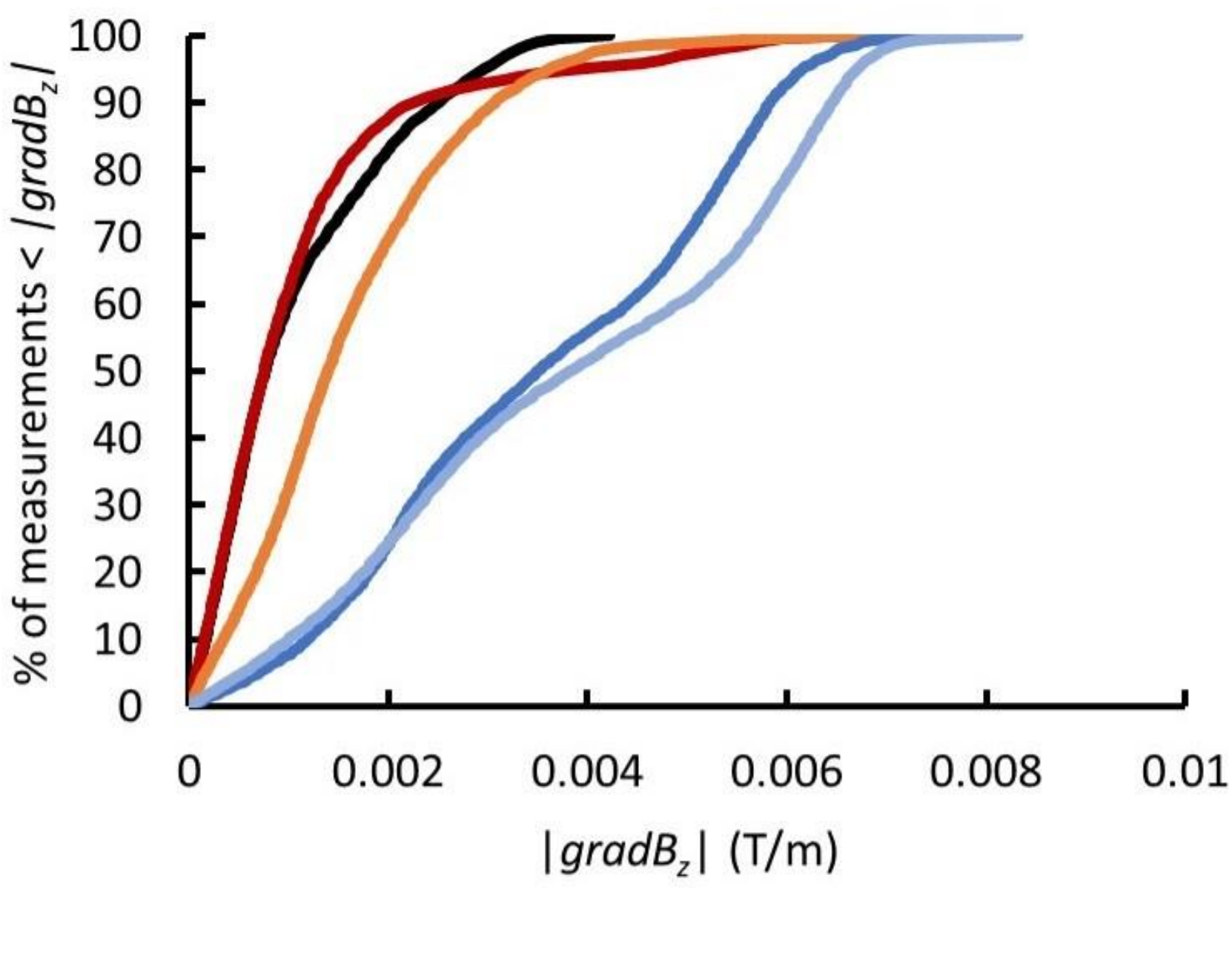

**Figure 5.** Cumulative distributions of $|gradB_z|$ measurements on the control plate for plate temperatures of 13 °C (light blue), 13 °C (blue), 38 °C (orange), 110 °C (red), and 166 °C (black).

The two scans of furnace 2 in Figure 4a and the two scans of the control plate at 13 °C in Figure 5 illustrate the reproducibility of the diagnostic method. The data at four temperatures show $|gradB_z|$ decreases with increasing temperature from 13 °C to ~38 °C and becomes insensitive to temperature above 110 °C.

Furnaces 1, 2, and 3 were shut down 19, 21.5, and 24 h, respectively, before their magnetic field profiles were scanned. The maximum temperature of the furnace bottoms during a heating cycle was measured to be ~180 °C. The measured temperatures of furnaces 1, 2, and 3 were the same during the magnetic-field scans within the error of the temperature measurements and were consistent with a cooling rate of 1.6 °C/hour or 150 °C, 146 °C, and 142 °C for furnaces 1, 2, and 3, respectively. Since all of these temperatures are well above the 110 °C at which the $|gradB_z|$ distribution is insensitive to temperature, as shown in Figure 5, we conclude that the scan at 166 °C adequately represents an MQN-free control to compare with the scans of the furnace bottoms.

To quantitatively understand how these data might relate to the MQN hypothesis, we simulated the $|gradB_z|$ distributions in 2.54 cm thick A36 steel as a function of MQN mass and areal number density.

### *3.2. Computer Simulations of MQNs in Steel Plates*

The COMSOL [14] Multiphysics finite-element code was used to calculate magnetic force and torque on one to 14 MQNs near and inside a 2.54 cm thick steel furnace bottom and a piece of magnetite ore. Details of the calculation and of the validation process are included in Appendix A, Method and validation of COMSOL Simulations.

Simulations of the vertical force on an MQN outside and inside a mild steel plate showed the magnetic force was towards the center of the plate and was maximum at the air-steel boundary. The force on an MQN inside the plate increased with increasing MQN mass, as shown in Figure 6. The force preventing an MQN from reaching the air-steel boundary increases from about 100 times the gravitational force for MQN mass of 0.1 kg to about 400 times gravity for 100 kg. The magnetic force on an MQNs inside the steel makes steel an excellent reservoir of massive MQNs but makes expulsion from the steel rods to the steel liners in a rod mill less likely for very large masses. Hence, the process is more likely to allow the low-mass MQNs (<1 kg) to populate the liner with the very few higher mass MQNs retained in the rods.

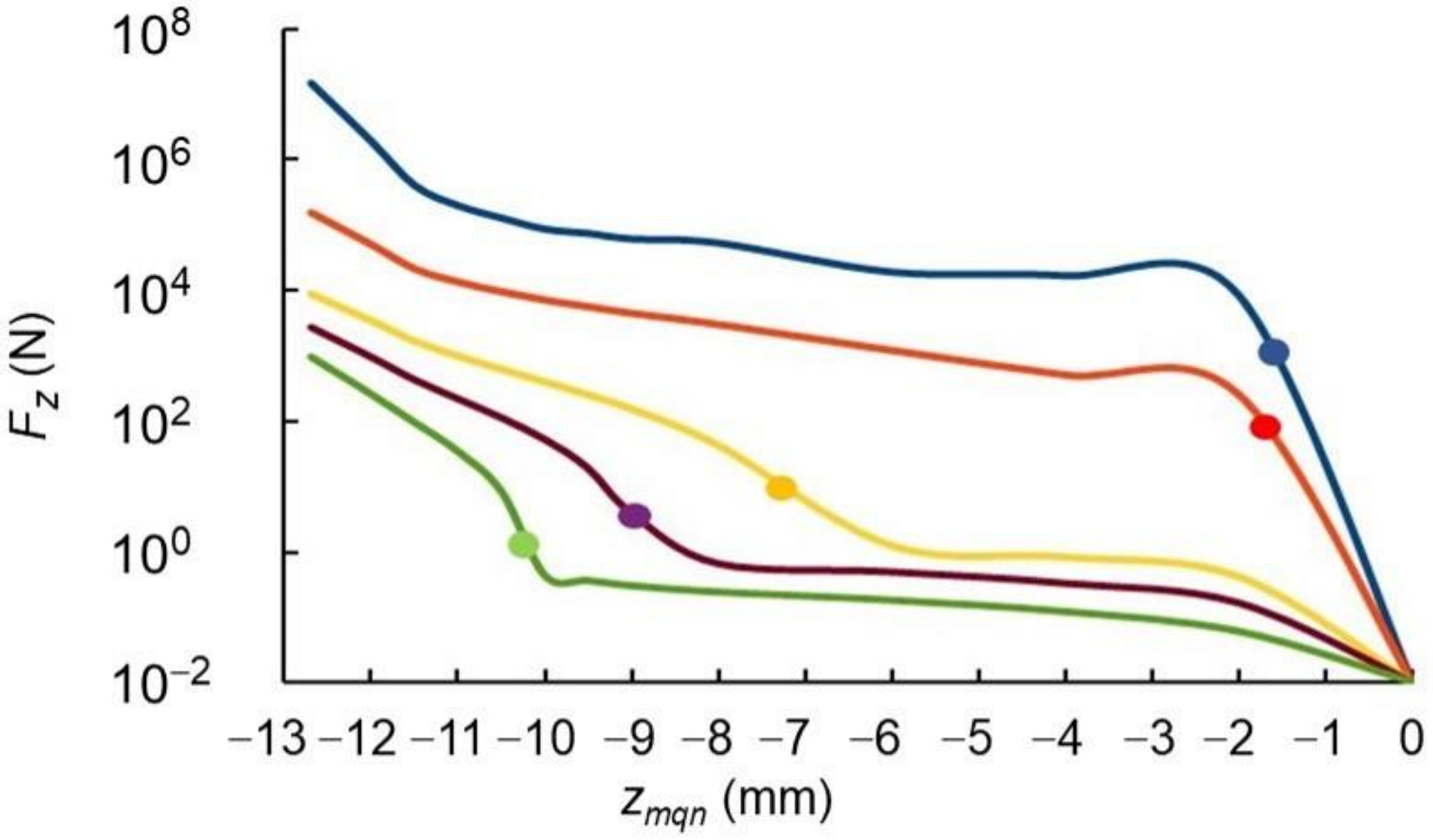

**Figure 6.** Upward magnetic force on an MQN inside a 25.4 mm thick steel plate as a function of distance from the plate center ($z_{mqn} = 0$) for MQN masses between 0.1 kg (light green), 0.32 kg (purple), 1.0 kg (yellow), 10 kg (red), and 100 kg (blue). The circles show the position at which the downward gravitational force is balanced by the upward magnetic force. The air–steel boundary is at $z_{mqn} = -12.7$ mm.

A similar calculation with the B-H curve of magnetite [23] instead of steel shows that the magnetic force on the MQN inside magnetite is much less than it is inside steel. For MQN mass of > 30 mg, the maximum magnetic force is greater than the gravitational force. Therefore, MQNs of mass > 30 mg should accumulate in magnetite ore.

The distance between the equilibrium position of the MQN, under magnetic and gravitational forces, and the center of the steel plate decreases with increasing MQN mass, as shown in Figure 6. Consequently, the larger magnetic field of larger MQNs is more shielded by the larger amount of steel between the MQN and the surface.

The shielding effect of steel between two 1.0 kg MQNs with their magnetic moments aligned and with both located at their equilibrium position 8.8 mm from the center of the 25.4 mm thick steel plate was assessed by calculating the attractive force as a function of separation d. The results are shown in Figure 7.

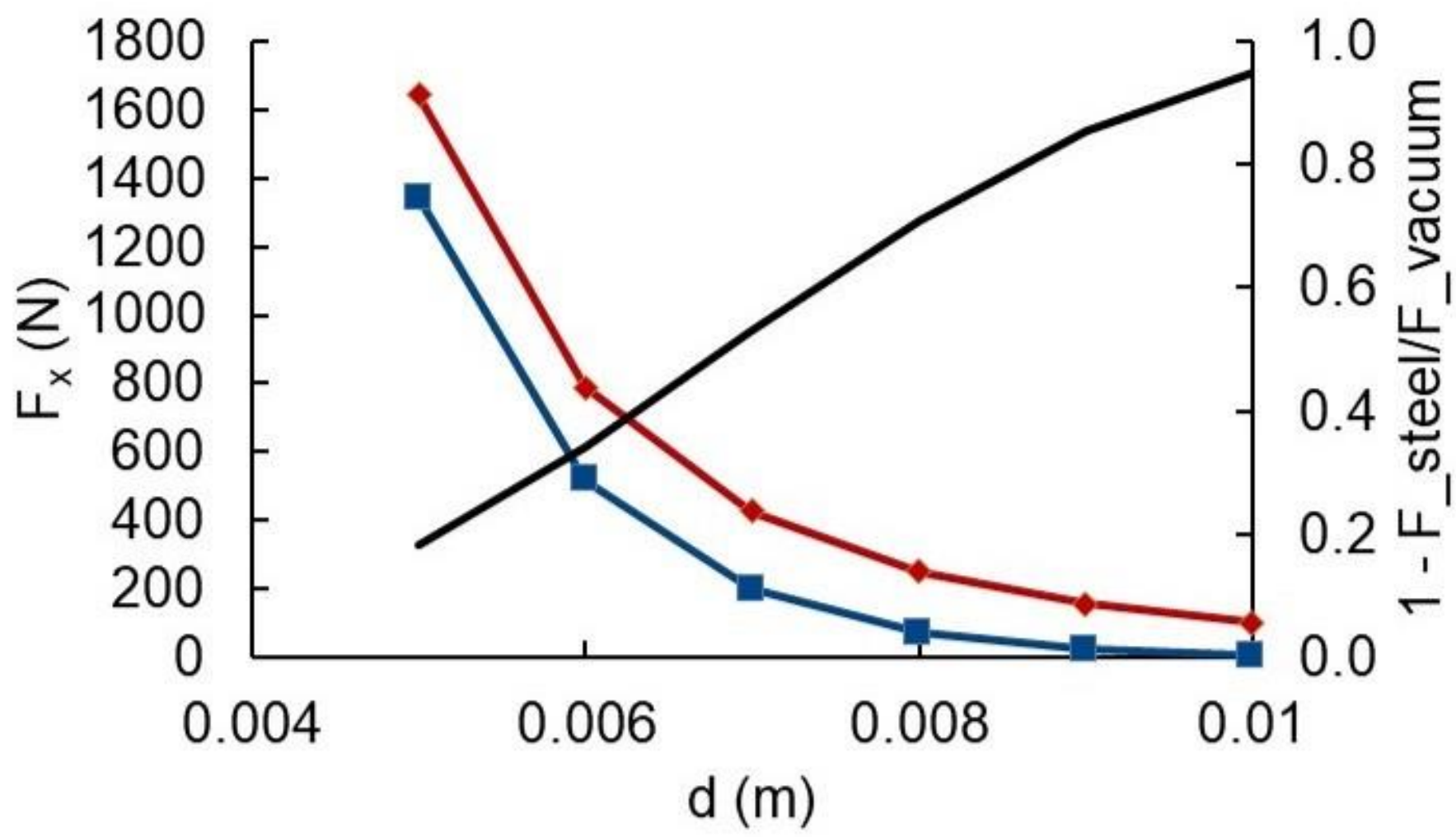

**Figure 7.** COMSOL simulation of the shielding effect of steel on the force between two 1 kg MQNs separated by distance *d* with their magnetic moments aligned. The attractive force $F_{steel}$ in steel (blue) and the attractive force $F_{vacuum}$ in vacuum (red) are shown on the left axis. The shielding coefficient (black) = 1 − $F_{steel}/F_{vacuum}$ and is shown on the right axis.

As shown in Figure 7, steel effectively shields an MQN. A separation of 7 mm reduces the attractive force between two 1 kg MQNs by 50% and a separation of 10 mm reduces the attractive force by 95%. Shielding by the steel and vector addition of the magnetic fields from multiple MQNs as a function of their separation *d* led to adoption of $|gradB_z|$ as the key diagnostic, as defined by Equation (5) and illustrated in Figure 2.

Furnace bottoms may contain multiple MQNs interacting with each other. Therefore, we simulated an array of 14 MQNs at their equilibrium position in 2.54 cm thick steel under two simplifying assumptions: (1) all MQNS are equidistant from their nearest MQN neighbors by length parameter *d*, and (2) all MQNs are assumed to have the same mass and magnetic moment.

We found the orientation of MQN magnetic moments materially affected the computed $|gradB_z|$ distributions. Since the accuracy of the simulations was insufficient to determine the pattern of magnet moments in an array, we explored four patterns, partially derived from laboratory emulations with magnetized spheres, and compared the results with the data shown in Figure 4a. The simulations and analysis are described in detail in Appendix D, Effect of magnetic-moment orientation on $|gradB_z|$ distributions. As shown in Figure A3, $|gradB_z|$ increases rapidly with decreasing *d* whenever *d* < 1.9 cm, as the coupling between the MQNs increases. Figure A2a shows that only the pattern from MQNs (1) with mass ~ 0.32 kg, (2) with separation *d* between 1.43 cm and 1.5 cm, and (3) with all magnetic moments aligned (Figure A1d) is consistent with the data in Figure 4a for furnaces 1 and 2, which have been in service for 41 years.

Unlike furnaces 1 and 2, all four patterns in Figure A1 generally agree with the data from furnace 3 after 9 years of operation, within the experimental variation inferred from the scans of furnaces 1 and 2 in Figure 4a. Figure A3 shows $|gradB_z|$ distributions are only weakly dependent on $d$ for 1.95 cm ≤ $d$ ≤ 3.0 cm, and Figure A2b shows that all four patterns of magnetic moment are similar for $d$ = 2.92 cm. Very similar results were obtained with 2.5 cm ≤ $d$ ≤3.5 cm. Insensitivity to $d$ and to pattern of magnetic moments is expected in this range of $d$ values, in which the $|gradB_z|$ diagnostic measures isolated MQNs since $d$ is much larger than the ~1 cm shielding distance.

The simulation with the pattern in Figure A1a (i.e., adjacent rows having opposite magnetic moments) gives the best fit to the furnace 3 data and is the only pattern, of the four tested, that matches the initial slope and the inflection point at $|grad\text{Bz}|$ = 0.002 T/m, as shown in Figure A2b.

Figure 8 compares the furnace data in solid lines and the best-fit COMSOL simulations assuming the $|gradB_z|$ below the furnace bottoms is caused by uniformly spaced MQNs of 0.32 kg mass.

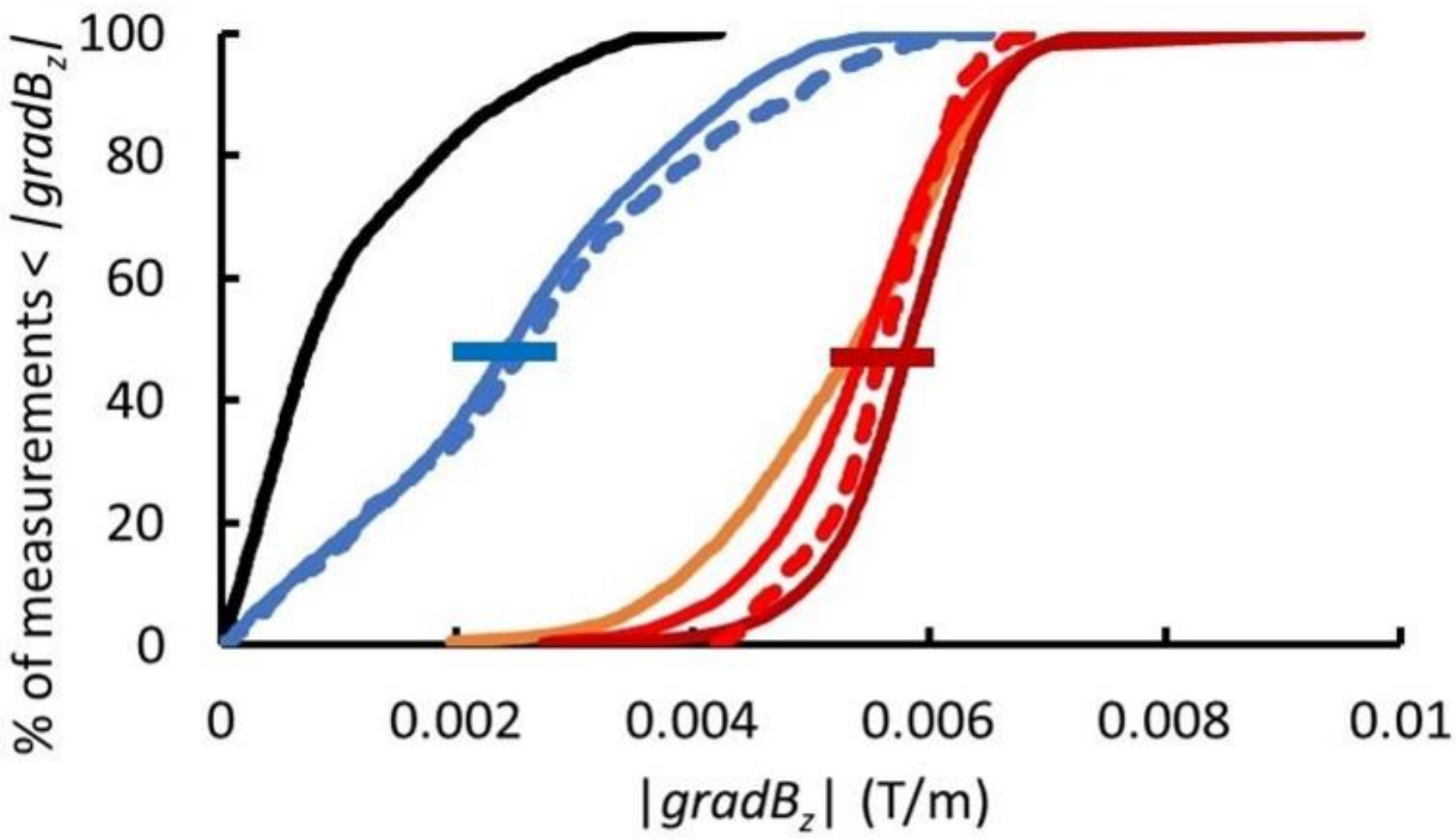

**Figure 8.** Cumulative distributions of $|gradB_z|$ measurements on the three furnaces and the control are shown: furnace 1 (orange), furnace 2 first scan (dark red), furnace 2 s scan (bright red), furnace 3 (blue), and control at 166 °C (black). Best-fit COMSOL simulation results are shown: (1) with MQN spacing d = 2.92 cm, MQN mass of 0.32 kg, and with MQNs' adjacent rows with opposite magnetic moments as shown in Figure A1a (dashed blue), and (2) with MQN spacing d = 1.46 cm, MQN mass of 0.32 kg, and with all MQN magnetic moments aligned, as shown in Figure A1d (dashed red). The different configurations of magnetic moments are tentatively attributed to the weak coupling for d = 2.92 cm and strong coupling for d = 1.46 cm since the shielding distance is ~1 cm, as shown in Figure 7. The significance the error bars is described in the caption to Figure 4.

## 4. Discussion

The distributions of $|gradB_z|$ from the furnace scans are consistent with simulations of MQNs from magnetite processing and are inconsistent with the corresponding scans of the steel control at the same temperature. However, the physics of centimeter-scale variations in the magnetic field perpendicular to a thick steel plate has not been well researched; therefore, there could be some unknown phenomenon consistent with the data and not requiring MQNs.

As shown in Figure 5, the distribution of $|gradB_z|$ with a steel-plate temperature of ~ 13 °C resembles those of furnace 3 after 9 years of operation. However, the temperature of furnace 3 was inconsistent with 13 °C. In addition, the distribution of $|gradB_z|$ in the control plate is not consistent with those of furnaces 1 and 2 after 41 years of operation. We conclude that the variation in the distribution of $|gradB_z|$ with temperature is not a viable alternative explanation of the results.

The agreement in Figure 4a between multiple scans of furnace 2 and between scans of furnace 1 and 2 (furnaces with the same 41 years of operation) similarly eliminates measurement variation as a potential alternative explanation.

The three furnaces were built by the same manufacturer (Whiting Equipment Canada) to the same design and with the same A36 steel and appear identical. However, there is the possibility that there was some unknown variation in the process that caused the different distribution of $|gradB_z|$ for furnace 3. There is no way of knowing for certain.

The dependences on MQN mass of (1) MQN number density, (2) acceptance factor, (3) magnetic force inside steel, and 4) eddy current force during MQN motion through steel, all preferentially select <1 kg MQNs for transfer into furnace liners. These dependencies and the insensitivity of the $|gradB_z|$ diagnostic for MQN mass < 0.1 kg means that our observations should be most sensitive to MQNs with mass between 0.1 kg and 1 kg, i.e., with 0.32 kg decadal mass. The data for furnaces 1 and 2 in Figure 8 are consistent with the simulations for single-mass 0.32 kg MQNs, with all MQN magnetic moments aligned and with a single inter-MQN spacing between 1.43 and 1.50 cm.

Simulations with (1) the full distribution of MQN masses, (2) some distribution of magnetic-moment orientations, and (3) some distribution of spacings between MQNs would be more realistic. However, (1) the dependence of magnetometer sensitivity and MQN stopping power on MQN mass means only a narrow range of decadal mass can cause the data from furnace scans, (2) simulations summarized in Figure A2 show the data from furnace 1 does not strongly depend on magnetic-moment orientation (since the large spacing indicates the MQNs are isolated) and show that the data from furnaces 1 and 2 agree only with one of the four magnetic moment orientations, and (3) single-spacing simulations summarized in Figure A3 show only a small range in spacing gives agreement with the data. Therefore, including a broader range of parameters may change the detailed shape of the simulated curves (potentially improving the agreement) but should not change the median points or the conclusions. In addition, there is no theory prescribing a preferred distribution in magnetic moment or spacing, so these more realistic simulations would have to be time dependent and allow these variables to change self-consistently under the computed forces and torques. Such simulations are currently beyond the state of the art.

If the $|gradB_z|$ below the furnace bottoms is caused by MQNs, then these results imply 1490 +/− 70 kg/m$^2$ of MQNs were accumulated during its 41 years of operation and about 200 +100/− 50 kg/y of MQNs accumulating per furnace with an effective area of ~6 +3/− 1.5 m$^2$ for the bottom. The large uncertainty in area and corresponding accumulation rate reflects the uncertainty in the MQN distribution outside the 4.5-m$^2$ scanned area.

Furnaces 1 and 2 have been operating for 41 years. Furnace 3 has been operating for 9 years. Plant throughput per furnace has been reasonably constant, so the area density of accumulated MQNs should be inversely proportional to the furnace age. The average inter-MQN spacing $d$ should be inversely proportional to the square root of furnace age, and the mean $|gradB_z|$ should be proportional to the square root of furnace age, as shown in Figure 4b.

The simulation results consistent with furnaces 1 and 2 (i.e., a single inter-MQN spacing of $d \approx 1.5$ cm with all MQN magnetic moments aligned) would imply that the spacing for furnace 3 should be ~3.0 cm. With such a large expected spacing, the MQNs should be independent of each other (as inferred from the shielding distance in Figure 7) and hence the orientation should be random. The average $|gradB_z|$ for the four simulated orientations shown in Figure A1, for MQN separation of 2.92 cm, is 0.0021 T/m with a standard deviation of 0.0004 T/m. This is compatible with the measured value of 0.0025 T/m for furnace 3. Hence the data from furnaces 1, 2 and 3 are compatible with a uniform accumulation rate of MQNs.

The data interpreted with the simulations indicate about 200 kg/y of MQNs accumulate per furnace if and only if the |$gradB_z$| below the furnace bottoms is caused by MQNs. At 4 tons per batch and 3 batches a day for 250 days per year, ME Elecmetal processes approximately 2700 metric tons of steel per furnace per year, including the 30% to 40% new steel to make up for liner wear. Therefore, each furnace would process approximately 1760 tons of liner material to accumulate about 200 kg of MQNs. The corresponding mean MQN concentration in used liner material is approximately 0.12 kg/ton of liner.

This inferred 0.12 MQN-kg/liner-ton concentration and historical data on magnetite processing leads to an estimate of the solar-MQN mass density if the |$gradB_z$| distributions are caused by MQNs. According to the official history of the Erie Mining Company (LTV site, now Polymet Mining Inc., Hoyt Lakes, MN, USA) [24], approximately 909 million tons of taconite were processed from 1957 to 2001 in about 30 parallel mill lines at the Erie plant. Therefore, each mill line processed about 0.69 million tons of taconite per year.

New liners weighed 121 metric tons per mill. They were replaced at approximately 18-month intervals, when about 40% of their mass had been eroded and their remaining mass was about 73 tons. Consequently, each set of mill liners were replaced after processing approximately 1.0 million metric tons of taconite in 1.5 years. If the |$gradB_z$| below the furnace bottoms is caused by MQNs, then

the concentration of MQN mass per ton of taconite ore =

(0.12 MQN-kg/liner-ton) × (73-liner-tons/1.0 × $10^6$-taconite-tons) =

$8.5 \times 10^{-6}$ MQN-kg/taconite-ton = $8.5 \times 10^{+3}$ MQN-kg/ore-gigaton.

Since the bio-chemical process that created the magnetite ended 1.8 Ga ago when the Sudbury meteorite impacted North America, the accumulation time is approximately 1.8 Ga. Therefore, the rate of MQN accumulation would be ~$4.7 \times 10^3$ MQN-kg/Gt/Ga.

The furnace scans are not sensitive to MQNs with mass < 0.1 kg and the acceptance fractions decline quickly with increasing mass above 1 kg, as shown in Table A4. Therefore, the decadal mass interval contributing to the furnace data should be ~0.3 kg, which is consistent with the furnace data as interpreted with the COMSOL simulations.

The above mass accumulation rate of ~$4.7 \times 10^3$ MQN-kg/Gt/Ga (for the first 200 m of ore deposit in a typical 200 m deep mine) is ~6000 times the value expected from interstellar dark matter which is 0.81 MQN-kg/Gt/Ga (from columns 1 and 2 of Table A3 for 0.32 kg MQN). It is also ~2000 times the solar mass accumulation rate of ~2.5 MQN-kg/Gt/Ga (from columns 1 and 2 of Table A5 for 0.32 kg MQN) unless the solar accumulation processes described above produce a mass enhancement of a factor of ~2000 for $Bo$ = 1.65 TT. Including the factor of ~5 uncertainty in the number density of 0.32 kg MQNs, still requires a factor of ~ 400 enhancement from the solar accumulation process. Both of these calculations assume that all dark matter would be composed of MQNs. However, the enhancement factor of the proposed solar-MQN accumulation process is currently unknown. If the enhancement factor is <<2000, then the data from the furnaces cannot be caused by MQNs or the simulation results [16] for the local density of dark matter must be in error. If the enhancement factor is >>2000 and if the data from the furnaces are determined to be caused by MQNs, then the fraction of dark matter attributable to MQNs must be <<1.

Even with the inferred factor of ~ 2000 mass enhancement from solar MQNs, the accumulated mass of all MQN dark matter in the center of the Earth during its ~4 Ga existence would only be ~ $5 \times 10^{15}$ kg ≈ $10^{-9}$ Earth masses. The radius of that MQN would be about 2 m and the magnetic field from the MQN would be about 0.0004 of Earth's current magnetic field. Therefore, the MQN hypothesis is consistent with available information about Earth's mass profile and surface magnetic field.

The characteristic lifetime of MQNs near Earth should be comparable to the ~ $3 \times 10^7$-y lifetime [18,19] of asteroids near Earth and well away from the Jupiter resonances near Mars orbit. Therefore, the theoretical processes for slowing down and transitioning interstellar MQNs from their hyperbolic orbits to elliptical orbits near Earth must have a

lossless trapping rate =

~$2 \times 10^{3.0+/-0.7}$ mass enhancement for solar MQN ×

$4 \times 10^{-22}$ kg/m$^3$ mass density of interstellar MQNs divided by
$3 \times 10^7$ y lifetime of near-Earth bodies in solar orbit ≈
$2.7 \times 10^{-26+/-0.7}$ kg m$^{-3}$ y$^{-1}$
to be consistent with the furnace data, as interpreted with COMSOL simulations, and with MQN hypothesis for dark matter.

Xu and Siegel [21] calculate a
lossless trapping rate ≈
~$10^{4+/-1}$ mass enhancement for solar MQN ×
$4 \times 10^{-22}$ kg/m$^3$ mass density of interstellar MQNs divided by
$4.5 \times 10^9$ y age of the solar system ≈
~$10^{-27+/-1}$ kg m$^{-3}$ y$^{-1}$.

Their upper limit (based only on gravitational scattering by the planets) is still a factor of ~3 below the mean required rate but is within the stated errors and is well within the uncertainty when the factor of 5 in the uncertainty of *Bo* is included. MQNs have the additional advantage of being slowed by their magnetopause interaction while passing through a small portion of the corona and chromosphere at grazing incidence and then scattering from planets to avoid returning to the Sun. Whether or not this additional process is sufficient to bring the inferred accumulation rate closer to the mean expected value has not been determined. However, even if the process is sufficient to support the enhancement factor for solar MQNs, determining that would still not be sufficient to definitively validate the MQN hypothesis.

The $|gradB_z|$ data and the COMSOL simulations summarized in Figure 8 are consistent with the MQN hypothesis, but these results are not incontrovertible since we cannot eliminate some unidentified cause of the $|gradB_z|$ distribution that does not involve MQNs. Extracting and isolating ~0.3 kg MQNs with nuclear-density and a ~1 TT magnetic field from rod-mill liners would be definitive. We attempted to isolate and extract MQNs from 52 rod-mill liners, each of which weighed approximately 600 kg, before they reach the furnaces at ME Elecmetal. The liner mass per year processed by the foundry and the inferred MQN-mass accumulation rate in the furnaces let us estimate that the 52 liners should contain approximately 6 MQNs. We suspended each liner at 45 degrees for at least 100 s so an MQN could flow through the steel to the bottom corner of the liner. That corner was then cut off with a plasma cutter while checking that the temperature of the bottom most edge of the corner remained ferromagnetic so any MQN would remain in the sample. The sample masses varied from 95 to 400 g, so the addition of a ~300 g MQN would be evident. None were found. The weights of each sample in air and in water were measured. The variation in the density was only ~1% and is consistent with the variation in the chemical composition from similar material. Our process depended on MQNs flowing through the steel under gravitational and eddy-current forces. If so, MQNs would fall through each vertically suspended liner to its bottom, which was cut off and processed for excess mass and for mass change upon heating to above the Curie temperature. No MQNs were detected. However, we subsequently found that hysteresis losses associated with MQN motion through the high-carbon steel of the liners produced a resistance to motion that was >8 times the gravitational force. MQNs could not flow through the liners to be concentrated and collected, so the null result did not invalidate the MQN hypothesis.

We considered intercepting liners before they reach the furnaces, suspending them from one end, inductively heating the top of the liner to a temperature much greater than the 770 °C Curie temperature, and slowly lowering the heater so that the hot zone progresses to the bottom of the liner. Above the Curie temperature, steel transitions from ferromagnetic to paramagnetic, so the magnetic force on an MQN should be greatly reduced in the hot zone. Since the gravitational force on an MQN divided by its cross-sectional area exceeds the yield strength of steel by many orders of magnitude, any MQN in the hot zone should fall to the bottom of the zone, travel with the zone to the bottom of the liner and fall from the liner into a low-mass steel collector. However, this process may fail because (1) the large magnetic-field of a 1/3rd kg MQN traps electrons in the conduction band to a distance of about 700 μm despite the disorienting effect of thermal agitation above the Curie temperature and (2) the lattice is severely disordered by the ~1.2-μm diameter, nuclear-density MQN, so the transfer of force from the trapped electrons to the lattice is not reliably predictable. Although this proposed process may work, failure to extract MQNs would not be a definitive determination that MQNs do not reside in the liners. Consequently, a null result would not falsify the MQN hypothesis.

Therefore, we propose that the only practical way to definitively test the MQN hypothesis is to secure an array of 2.54-cm-thick, mild-steel (e.g., A36) plates to the top surface of a furnace bottom when a furnace is being refurbished, letting the plates collect MQNs released from the melted liners (instead of their reaching the furnace bottom), and removing the plates at the next furnace refurbishment. Each plate would be weighed before placement and after recovery. A mass increase consistent with the ~200 kg/y/furnace accumulation rate would be definitive. In addition, inductively heating the plates to melt would let the MQNs be extracted into a lower-mass steel collector or into a superconducting bowl for confirmation of nuclear density and ~1 TT magnetic field.

## 5. Conclusions

The measured distributions of $|gradB_z|$ from the three furnaces and a control sample are found to be consistent with uniform accumulation with time of MQNs from magnetite processing. Simulations are able to reproduce the observed effect. The best agreement of data and simulations are consistent with the theoretical mass distribution of MQNs and with the mass dependencies in acceptance factors in rock and transfer efficiencies in ore processing. Consistency does not imply proof, of course, since there may be other, yet unidentified, causes of the same $|gradB_z|$ distributions. If the measured $|gradB_z|$ distributions are caused by MQNs, then the data and simulations indicate ~200 kg of approximately 0.3 kg MQNs accumulate per year per furnace. The corresponding accumulation rate is consistent with ~1 kg of ~0.32 kg MQNs per $1.2 \times 10^8$ kg of taconite ore processed.

Even if the $|gradB_z|$ below the furnace bottoms is caused by MQNs, consistency with dark matter requires MQNs to have accumulated in the solar system to a mass density of ~$2 \times 10^{3.0+/-0.7}$ (i.e., ~400 to ~10,000) times the mass density of interstellar dark matter by a yet-to-be quantified combination of (1) purely gravitational scattering with the planets and (2) first slowing down during passage through a portion of the Sun's corona and/or chromosphere at grazing incidence and subsequently scattering by planets into near-Earth orbits. Calculating the net density of MQNs accumulating near Earth through these processes is necessary but not sufficient to connect the furnace data with the MQN hypothesis for dark matter.

Deploying one or more collector plates above a furnace bottom (at ME Elecmetal or an equivalent foundry processing rod-mill liners from magnetite ore processing) for one refurbishment cycle and measuring a mass increase consistent with the inferred MQN accumulation rate would be a definitive test of the MQN hypothesis.

**Author Contributions:** Conceptualization, J.P.V., T.S. and M.G.; methodology, J.P.V.; software, J.P.V. and T.S.; validation, J.P.V., T.S. and M.G.; formal analysis, J.P.V. and T.S.; investigation, J.P.V.; resources, J.P.V.; data curation, J.P.V., T.S. and M.G.; writing—original draft preparation, J.P.V.; writing—review and editing, J.P.V., T.S. and M.G.; visualization, J.P.V. and T.S.; supervision, J.P.V.; project administration, J.P.V.; funding acquisition, J.P.V. All authors have read and agreed to the published version of the manuscript.

**Funding:** This work was funded entirely by VanDevender Enterprises, LLC, Albuquerque, NM, USA under grant 20211210.

**Data Availability Statement:** The datasets generated, and the computer programs developed, during the current study are available from the corresponding author on reasonable request.

**Acknowledgments:** We gratefully acknowledge (1) many suggestions and practical insights about taconite geology, mining and processing by Andrew Ware, Dave Hughes, and James Tieberg of PolyMet Corp, and by Dean Peterson of Big Rock Exploration,; (2) technical insights on the solid state physics of MQNs embedded in steel from Paul A. Fleury and on the engineering aspects of electric arc furnaces from Keith Tolk, John Boyes, and Ron Parriott; and (3) expert assistance with attempts to extract MQNs from liners by Tom Storks and Criss Swaim and by James Bougalis, John Bougalis, Tiffany King, Johann Grobler, Gary Liubakka, and staff at Scranton Iron Works.

**Conflicts of Interest:** Author J. Pace VanDevender is the President of VanDevender Enterprises, LLC, which is a for-profit company supporting energy research and development through technical consulting services, and Principal Investigator of the MQN Collaboration, which is an international collaboration of volunteers exploring the Magnetized Quark Nugget (MQN) hypothesis for dark matter. Author Michael Glissman is employed by the company Polymet Mining Inc. The authors declare that this work was funded entirely by VanDevender Enterprises. In addition, the authors have not received nor expect to receive any compensation for their participation in this paper. Therefore, the authors declare that the research was conducted in the absence of any commercial or financial relationships that could be construed as a potential conflict of interest.

## Appendix A. Method and Validation of COMSOL Simulations

The commercially available COMSOL [14] Multiphysics finite element code with the Magnetic Fields No Current (MFNC) option [25] in the AC/DC Module efficiently solves the magnetic flux conservation equation for the magnetic scalar potential over a three-dimensional mesh, with the mesh resolution defined by the physics of the problem. The application was used in the magnetostatic mode to calculate magnetic force and torque on an MQN inside a piece of magnetite ore, and on one to 14 MQNs near or inside a 2.54 cm thick steel furnace bottom.

COMSOL models a continuous and homogeneous ferromagnetic material in three dimensions with an embedded sphere of specified magnetic moment simulating an MQN. Microscale physics that might arise from a massive, micron-sized MQN with trillion tesla magnetic field is not yet understood and is not modeled. Therefore, the results can only guide the experiment, not replace it. This is one reason the consistency of the measurements with the simulations of MQNs in steel is not dispositive. Only extraction of a massive, nuclear-density MQN from the material will be definitive.

The MQN was modeled as a sphere of uniform magnetization that had the same magnetic moment as the MQN with a given radius (determined by its mass and nuclear mass density) and with a surface magnetic field $B_o$ on the equator. The value of $B_o$ = 1.65 TT +/− 0.35 TT, as inferred in References [9,10,12]. To avoid overstressing the meshing capability of the code, the sub-micron diameter of MQNs was approximated by a 500-micron radius magnet with the same magnetic moment of the MQN. Runs maintaining a constant magnetic moment of the MQN and varying the radius of the MQN between 0.5 microns and 500 microns gave the same force (within ~26%) on the MQN at a fixed position inside the steel plate.

The 500 μm sized simulated MQN is the minimum size for the code to reliably converge and provide the force and torque on the MQN. At smaller scales the mesh generated pseudo force calculated with relative permeability of 1 is larger than the force computed with the same mesh but with the ferromagnetic material. Varying the size and extrapolating the force and torque to 1 micron scale gave us confidence that the results were sufficient to support the conclusions.

The extremely high magnetic field of MQNs was accommodated by extending the interpolation table of the included B-H curve of "Soft Iron (Without Losses) to $B$ = 2 TT at $H = 1.59 \times 10^{18}$ A/m. The extension ensured that the space immediately around the MQN and extending to where $B \sim 2.4$ T at $H \sim 3.5 \times 10^5$ A/m is magnetically saturated. MQNs with the same magnetic moment and sufficiently small to saturate the same volume of space around it will generate comparable magnetic fields in the steel and air.

The same method was used to explore the magnetic force holding an MQN inside a piece of magnetite ore. For these calculations, the B-H curve in the interpolation table was multiplied by 0.215 for $0 \leq B \leq 0.35$ T and extended from $B$ = 0.35 T at $H$ = 6300 A-turn/m at constant slope $\delta B/\delta H = \mu_o$, the permeability of free space, for 0.35 T $\leq B \leq$ 2.0 TT.

The problem of an MQN at equilibrium in a steel plate and under the combined gravitational and magnetic forces is necessarily not symmetric about any point. Therefore, the computational mesh is not symmetric and creates a pseudoforce on the MQN. However, we found that the net force $F_{net} = F_{BH} - F_{\mu r1}$, in which $F_{BH}$ is the force computed with the magnetic properties of the steel given by the B-H curve and $F_{\mu r1}$ is the force computed with those magnetic properties given by the relative permeability set to 1 (through the Magnetic Flux Conservation setting).

The procedure for calculating the force with $F_{net}$ was validated by experiments. The calculated force agrees with the observed force attracting a spherical neodymium magnet to a slab of steel within the 10% uncertainty of the sphere's magnetization. The same procedure gives the force and torque on the magnet as a function of depth of the magnet in a hole in a 2.54 cm thick iron slab. Friction with the walls of the hole prevented quantitative measurements. However, the results agree qualitatively with the observed equilibrium position (center of slab) and orientation (magnetic moment perpendicular to the hole axis).

Detecting MQNs inside a steel furnace bottom is complicated by magnetic domains in the steel shielding the magnetic field of an embedded MQN and by the ~7.5 mm gap between the steel and the magnetometer required to protect the magnetometer from the ~180°C temperature of the steel bottom during furnace operation. The distance between the equilibrium position on an MQN and the steel-air interface increases with increasing MQN mass, as shown in Figure 6. However, the MQN magnetic field also increases with increasing mass and saturates the steel to a larger distance from the MQN. The net effect is that the shielding effect decreases with increasing mass. The ratio of peak $B_z$ divided by the magnetic field that would be there in vacuum $|B_z/B_{vac}|$ at the sensor located at 7.5 +/− 1 mm from the steel-air surface is

$$\left|\frac{B_z}{B_{vac}}\right| \approx 0.06\, m_{mqn}^{0.92}, \tag{A1}$$

which varies from 0.009 to 0.53 for MQN mass $m_{mqn}$ between 0.1 kg and 10 kg, respectively.

The maximum value of $B_z$ at the sensor is

$$|B_z| \approx 0.0034\, m_{mqn}^{1.62}, \tag{A2}$$

which varies from 0.0001 T to 0.15 T for MQN mass $m_{mqn}$ between 0.1 kg and 10 kg, respectively. For $m_{mqn} < 0.0001$ kg, Earth's magnetic field and variations in the magnetic domains in the steel preclude observing $B_z$ from the MQN. The spatial variation in both of these background fields occur on a larger scale length than the $B_z$ from the MQN, so $|gradB_z|$ provides a more reliable diagnostic for MQNs. The COMSOL simulations give

$$|gradB_z| \approx 0.35\, m_{mqn}^{1.645}, \tag{A3}$$

which varies from 0.0086 T/m to 13.5 T/m for MQN mass $m_{mqn}$ between 0.1 kg and 10 kg, respectively. We find that a two-magnetometer sensor (with magnetometers separated by ~ 1.0 cm) and positioned on a rolling platform (to maintain a steel-to-magnetometer distance of 7.5 +/− 1 mm) to measure the gradient $|gradB_z|$ can reliably detect MQNs of mass > 0.1 kg.

## Appendix B. Calculation of Acceptance Fractions

MQNs reach Earth by passing through the ionized matter of Earth's magnetosphere and ionosphere. That plasma is heated by stagnation against the MQN magnetosphere. We assume that radiation from the heated plasma propagates upstream from the MQN to ionize the matter in the troposphere and land to sustain the magnetopause. At low velocities, the extreme magnetic field of the MQN is sufficient to ionize matter for atoms in which the energy imparted to electrons from Zeeman splitting of the ground-state energy level exceeds the ionization energy. We are not currently capable of doing self-consistent calculations of this ionization process, so the justification for this assumption rests on the observation that reentry of spacecraft at much lower velocity ionizes the atmosphere in front of the spacecraft and on observations [9,10] that depend on this assumption.

For an MQN of mass density $\rho_{mqn}$ and mass $m_{mqn}$, velocity $v_{mqn}$ moving through material of density $\rho_p$ with a drag coefficient $K \approx 1$, and surface magnetic field at the equator $B_S$, the geometric radius of the MQN is

$$R_{mqn} = \left(\frac{3\, m_{mqn}}{4\pi\rho_{mqn}}\right)^{1/3}, \quad \text{(A4)}$$

and the plasma-facing surface of the approximately spherical magnetopause has a radius $R_{mp}$ given [15] by

$$R_{mp} \approx \left(\frac{2B_S^2}{\mu_0\rho_p v_{mqn}^2}\right)^{\frac{1}{6}} R_{mqn}, \quad \text{(A5)}$$

in which $\mu_o$ is the permeability of free space. The cross section for momentum transfer is given by the area of the MQN magnetopause and the stopping power is given by

$$\frac{dE}{dx} = \pi R_{mp}^2 v_{mqn}^2 \rho_p. \quad \text{(A6)}$$

In this form, the stopping power depends explicitly on the unknown surface magnetic field $B_S$ and implicitly on the unknown MQN mass density $\rho_{mqn}$. However, the Monte Carlo simulation [6] that produced the MQN mass distribution as a function of the magnetic field parameter $B_o$ showed that $B_S$, $\rho_{mqn}$, and $B_o$ are related by

$$< B_S >= \left(\frac{\rho_{mqn}}{10^{18}(kg/m^3)}\right)\left(\frac{\rho_{DM_T=100Mev}}{1.6\cdot 10^8(kg/m^3)}\right) B_o = k\, \rho_{mqn}\, B_o, \quad \text{(A7)}$$

which defines $k$ as

$$k = 10^{-18}\left(\frac{\rho_{DM_T=100Mev}}{1.6\cdot 10^8(kg/m^3)}\right). \quad \text{(A8)}$$

For $\rho_{DM_T=100MeV} = 1.6 \times 10^8$ kg/m$^3$ = the density of dark matter at time ~65 μs, when the temperature of the Universe was ~100 MeV, $k = 10^{-18}$.

Combining Equations (A4), (A5) and (A7) gives an expression for $R_{mp}$ that is independent of $\rho_{mqn}$:

$$R_{mp} = \left(\frac{9\, k^2\, B_o^2\, m_{mqn}^2}{8\,\pi^2\,\mu_0\,\rho_p\, v_{mqn}^2}\right)^{\frac{1}{6}}. \quad \text{(A9)}$$

Observations [9,10] limited the $B_o$ parameter to $B_o$ = 1.65 +/− 0.35 TT. The $B_o$ parameter characterizes the MQN mass distribution which was computed [6] only by decadal mass interval, defined as all MQN masses with the same value of the function Integer(($\log_{10}$ (mass)). Each decadal interval is represented by the logarithmic mean of the interval, i.e., $\sqrt{10}$ times the minimum mass in the interval. Assuming the interstellar MQN mass density for all MQN masses is the mass density of the local dark matter halo = 4 × $10^{-22}$ kg/m$^3$ [16,17], the number of MQNs m$^{-3}$ in interstellar space is given in Table A1 for the lower, middle, and upper values of $B_o$ and for the 0.3 kg decadal mass interval of most interest for this paper. Table A1 provides a wider decadal-mass range (0.003 kg to 300 kg) to help assess other iron ore deposits.

**Table A1.** Calculated MQN number density in interstellar space as a function of $B_o$ parameter and MQN decadal mass interval assuming the total MQN mass density equals the local mass density of the dark matter halo.

| Approximate Decadal Mass $m_{dec}$ (kg) | Interstellar MQNs/m³ in Decadal Mass Interval for $B_o$ = 1.3 TT | Interstellar MQNs/m³ in Decadal Mass Interval for $B_o$ = 1.65 TT | Interstellar MQNs/m³ in Decadal Mass Interval for $B_o$ = 2.0 TT | Mean and Variation in $Log_{10}$ of Number Density in Units of MQNs/m³ |
|---|---|---|---|---|
| 0.003 | $5.0 \times 10^{-26}$ | $1.1 \times 10^{-26}$ | $2.8 \times 10^{-27}$ | −25.97 +0.66/−0.58 |
| 0.03 | $1.9 \times 10^{-26}$ | $5.2 \times 10^{-27}$ | $1.5 \times 10^{-27}$ | −26.28 +0.57/−0.55 |
| 0.3 | $3.0 \times 10^{-26}$ | $6.3 \times 10^{-27}$ | $1.5 \times 10^{-27}$ | −26.20 +0.68/−0.61 |
| 3 | $2.0 \times 10^{-26}$ | $5.0 \times 10^{-27}$ | $1.4 \times 10^{-27}$ | −26.30 +0.59/−0.57 |
| 30 | $3.6 \times 10^{-27}$ | $1.3 \times 10^{-27}$ | $3.6 \times 10^{-28}$ | −26.87 +0.43/−0.57 |
| 300 | $1.6 \times 10^{-27}$ | $7.6 \times 10^{-28}$ | $1.7 \times 10^{-28}$ | −27.12 +0.33/−0.65 |

The uncertainty in the MQN number density arising from the uncertainty in the value of $B_o$ is represented in the last column of Table A1 as the variation in the $Log_{10}$ of the MQN number density. The uncertainties are approximately a factor of ~4.8 larger or ~1/4th as large as the value at $B_o$ = 1.65 TT.

The fraction of the MQNs stopping in rock was computed as follows. The dark matter is assumed to be composed of discrete MQNs with a velocity distribution [16] computed for Cold Dark Matter, including MQN dark matter. The velocity distribution has a random velocity, (approximated by a Maxwell-Boltzmann distribution with most probable velocity of 220 km/s. Superimposed on this random velocity is a streaming velocity of 230 km/sec, which is the velocity of the Solar System about the galactic center. The convolution integrals of this velocity distribution with the energy losses described by Equations (A6), (A8) and (A9) as the MQNs pass though the atmosphere and rock were computed numerically by the Monte Carlo technique. Full details of these integrals are given in [12]. The fraction of MQNs of interstellar origin (assuming a streaming velocity of ~230 km/s and "thermal velocity of ~220 km/s) with $B_o$ = 1.65 TT stopped in 100 m increments are shown as a function of MQN decadal mass in Table A2.

**Table A2.** Computed acceptance fractions of interstellar MQNs stopping in rock of density 2700 kg/m³ for $B_o$ = 1.65 TT as a function of MQN decadal mass and for depth intervals of 100 m.

| Approximate Decadal Mass $m_{dec}$ (kg) | Depth Interval 0 to 100 m | Depth Interval 100 to 200 m | Depth Interval 200 to 300 m | Depth Interval 300 to 400 m | Depth Interval 400 to 500 m |
|---|---|---|---|---|---|
| 0.003 | 0.27392 | 0.4234 | 0.22934 | 0.0294158 | 0.00041393 |
| 0.03 | 0.051 | 0.0927 | 0.152 | 0.189 | 0.208 |
| 0.3 | 0.00999 | 0.0221 | 0.031 | 0.0421 | 0.0471 |
| 3 | 0.00227 | 0.00341 | 0.00656 | 0.00874 | 0.01 |
| 30 | 0.000433 | 0.000664 | 0.0016 | 0.00137 | 0.00196 |

The results shown in Tables A1 and A2 are combined, as described in the main paper, to give the mass accumulation rate of interstellar MQNs stopped in taconite or magnetite. The accumulation rates for $B_o$ = 1.65 TT are shown in Table A3.

**Table A3.** Estimated $\frac{dF_{mqn}}{dt}$ = rate of mass accumulation of interstellar MQNs in a taconite or magnetite mine for five decadal mass intervals in units of kg mass of MQNs per gigaton of rock per Ga accumulation time, assuming the MQN speed $v_{mqn}$ = 230 km/s for all MQNs. The fraction that is stopped in the air is included in the 0 to 100 m interval since these would drift to the surface and fall into first magnetite layer.

| Approximate Decadal Mass $m_{dec}$ (kg) | Depth Interval 0 to 100 m | Depth Interval 100 to 200 m | Depth Interval 200 to 300 m | Depth Interval 300 to 400 m | Depth Interval 400 to 500 m |
|---|---|---|---|---|---|
| 0.003 | 1.0 | 0 | 0 | 0 | 0 |
| 0.03 | 0.979 | 0.021 | 0 | 0 | 0 |
| 0.3 | 0.33 | 0.454 | 0.157 | 0.0337 | 0.0008 |
| 3 | 0.093 | 0.100 | 0.200 | 0.183 | 0.106 |
| 30 | 0.016 | 0.025 | 0.040 | 0.054 | 0.051 |

Solar MQNs are interstellar MQNs trapped in the solar system by their interaction with the Sun and planets have a velocity of ~10% of the velocity of interstellar MQNs. We assume the mass distributions of Table A1 apply to solar MQNs and compute their acceptance fractions with the same Monte Carlo technique applied to interstellar MQNs. The results are shown in Table A4.

**Table A4.** Computed acceptance fractions of solar MQNs stopping in rock of density 2700 kg/m$^3$ for $B_o$ = 1.65 TT as a function of MQN decadal mass and for depth intervals of 100 m.

| Approximate Decadal Mass $m_{dec}$ (kg) | Depth Interval 0 to 100 m | Depth Interval 100 to 200 m | Depth Interval 200 to 300 m | Depth Interval 300 to 400 m | Depth Interval 400 to 500 m |
|---|---|---|---|---|---|
| 0.003 | 0.999 | 0.00 | 0.00 | 0.00 | 0.00 |
| 0.03 | 0.979 | 0.02 | 0.00 | 0.00 | 0.00 |
| 0.3 | 0.331 | 0.45 | 0.16 | 0.03 | 0.00 |
| 3 | 0.096 | 0.10 | 0.20 | 0.18 | 0.00 |
| 30 | 0.016 | 0.025 | 0.04 | 0.054 | 0.052 |
| 300 | 0.003 | 0.006 | 0.009 | 0.011 | 0.011 |

As was performed above for interstellar MQNs in Table A3, the acceptance fractions for solar MQNs and the MQN mass distribution are combined with Equation (2) to estimate the accumulated number of solar MQNs per unit volume of taconite ore per Ga of accumulation time as a function of decadal MQN mass—if the number density of solar MQNs is the same as it is for interstellar MQNs in Table A1, i.e., we assume an enhancement factor of only 1 for this calculation. The results are shown in Table A5.

**Table A5.** Estimated $\frac{dF_{mqn}}{dt}$ = rate of mass accumulation of solar MQNs in a taconite or magnetite mine for five decadal mass intervals in units of kg of decadal mass of MQNs per gigaton of rock per Ga accumulation time, assuming the MQN number density in the solar system is the same as it is in interstellar space and the speed $v_{mqn}$ = 30 km/s for all MQNs.

| Approximate Decadal Mass $m_{dec}$ (kg) | Depth Interval 0 to 100 m | Depth Interval 100 to 200 m | Depth Interval 200 to 300 m | Depth Interval 300 to 400 m | Depth Interval 400 to 500 m |
|---|---|---|---|---|---|
| 0.003 | 0.06 | 0.00 | 0.00 | 0.00 | 0.00 |
| 0.03 | 0.27 | 0.01 | 0.00 | 0.00 | 0.00 |
| 0.3 | 1.1 | 1.5 | 0.52 | 0.11 | 0.00 |
| 3 | 2.5 | 2.6 | 5.3 | 4.81 | 0.00 |
| 30 | 1.1 | 30. | 24. | 4.7 | 3.6 |
| 300 | 1.1 | 2.3 | 3.4 | 4.5 | 4.5 |

The lower velocity of solar MQNs gives approximately the same number density accumulation rate (ratio of mass accumulation rate divided by decadal mass) of 30 kg and 3 kg MQNs for 0 to 200 m and that rate is about 20% of the rate for 0.3 kg MQNs. However, the dependences on MQN mass of (1) MQN number density, (2) acceptance factor, (3) magnetic force inside steel, and 4) eddy current force during MQN motion through steel, all preferentially select low-mass (<1 kg) MQNs for transfer into furnace liners.

## Appendix C. Holding Force of Steel and Magnetite on Embedded MQNs

As shown in Figure 6, COMSOL simulations show the magnitude of the static magnetic force on an MQN inside a 2.54 cm thick steel slab is directed towards the center of the slab at $z_{mqn} = 0$, increases with distance from the slab center, and increases with increasing MQN mass. The distance between the steel–air interface (at $z_{mqn} = -12.7$ mm) and the equilibrium position (under gravitational and magnetic forces) increases with increasing mass. The ratio of magnetic force to gravitational force near the air-steel interface also increases with increasing MQN mass. Consequently, steel components of magnetite processing equipment may accumulate MQNs through static forces for sufficiently large MQN mass. Simulations show the magnetic attraction of MQNs to magnetite is greater than the gravitational attraction of the MQN to the Earth if the MQN mass is ≥approximately 30 mg.

However, ore processing is not static. MQNs may be ejected from magnetite by explosive-driven motion, by dropping ore from too large a height on the way to the mills, or by attraction to high-permeability steel rods (in rod mills) as magnetite grain size is reduced by the grinding process. Steel rod-mill rods are thrown across the mill at high speed into steel liners. MQNs accumulated in steel rods may gain enough speed relative to the steel to overcome the retaining force shown in Figure 6 and the eddy-current force (arising from the motion of MQNs through steel with measured conductivity of $9.3 \times 10^6$ S/m +/− 15%) and the hysteresis force (defined by the hysteresis losses in the ferromagnetic material per unit distance the MQN moves through the material) transfer from steel rods to steel liners upon impact. Since each mill has its own and often proprietary process, sufficient information to calculate the probability of accumulation of MQNs in a mill's liners is not usually available. In addition, the COMSOL computer code understandably has some difficulty accurately calculating forces on MQNs.

The potential transfer of MQNs from the rods to the liners is too complex and has too many unknows to be calculable. If MQNs exist, they are transferred to the liners and the rods as the magnetite is pulverized. We looked for MQNs in rods by measuring the mass density of 30 pieces of rods (with all corrosion removed) that were worn to their end of life. The 30 pieces weighed a total of 46 kg and represented 8.8 m of rod. Estimating the ratio of rod-rod impacts compared to rod-liner impacts and assuming the data from the furnaces at ME Elecmetal are indeed caused by MQNs from liners, we would expect 3 to 9 kg of MQNs (a 6% to 20% excess mass density) in the 30 samples. None of them had an extra 0.1 to 1.0 kg of mass above the mass expected for the measured density of 7825 kg $m^{-3}$ +/− 0.3%. The results implied that either the furnace data are caused by some phenomena not related to MQNs or the great majority of MQNs are transferred to the liners.

## Appendix D. Effect of Magnetic-Moment Orientation on |$gradB_z$| Distributions

We could not determine from first principles the pattern of MQN magnetic moments within the steel furnace bottoms. Therefore, COMSOL simulations calculated |$gradB_z$| distributions that would be measured if the MQNs had each of four patterns of magnetic moments within the x-y plane at the equilibrium distance $z$ from the bottom surface of a 2.54 cm thick steel plate. The four patterns are shown in Figure A1. The resulting |$gradB_z$| distributions were compared to the data shown in Figure 4a.

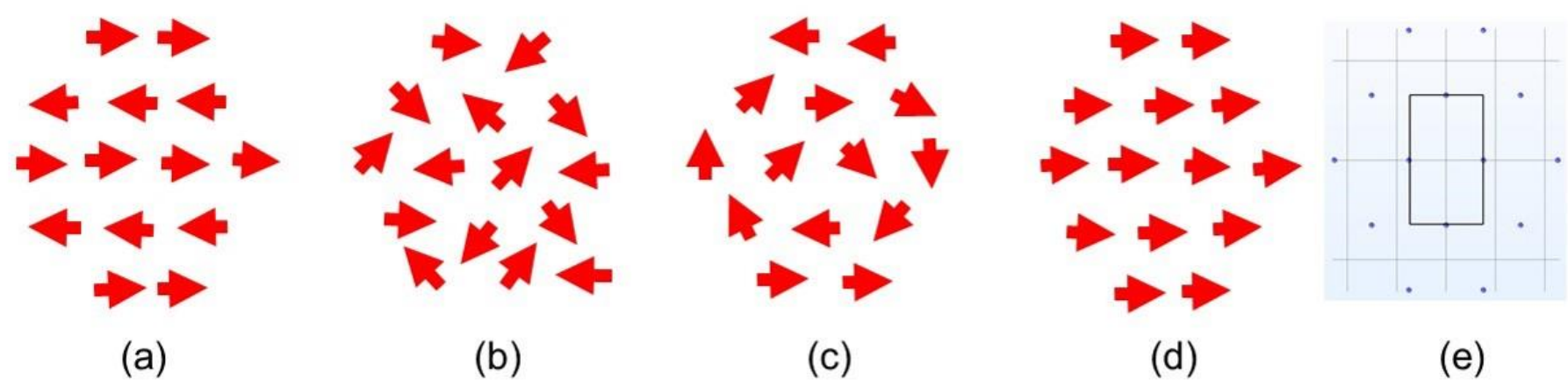

**Figure A1.** Patterns of the x-y orientation (shown as arrows) of magnetic moments examined with simulations: (**a**) adjacent rows have opposite magnetic moments, (**b**) clusters of three MQNs in a cell, (**c**) emulation in non-magnetic lattice of 14 neodymium magnetic spheres, and (**d**) all magnetic moments are in the same (+*x*) direction. (**e**) shows the pattern of the 14 MQNs (dots) with the area in the rectangle over which values for $B_1$ are computed in COMSOL simulations discussed below. $B_2$ is recorded at 1 cm to the right of the position of $B_1$. MQNs outside the scanned area contribute to the field as next nearest neighbors. The results approximate the scan of a much larger area with the same pattern.

COMSOL calculations of the total magnetic energy and the torque on a representative MQN are complicated by the problems of (1) the lack of a technique to compensate for mesh-asymmetry effects with multiple MQNs in a simulation and (2) the error introduced by taking the difference between two large numbers with their inherent computational uncertainty. Therefore, the simulations cannot unambiguously determine a clearly preferred pattern of magnetic moments in a steel medium. In addition to the emulation with a non-magnetic (wood) lattice shown in Figure A1c, an emulation was run with a ferromagnetic (steel) lattice to obtain some insight on the effect of immersion in steel. Fourteen 12.7 mm diameter neodymium magnets were inserted into 14 holes in a 6.35 mm thick steel plate with 30 mm separation between hole centers. The 14-hole pattern is the same as shown in Figure A1. The orientation of the magnetic moments was random. In the first trial, 12 of the 14 magnets self-aligned in the same direction, which is most consistent with all magnetic moments aligned, as shown in Figure A1d. Repeating the emulation a second time with the residual magnetic field of the steel from the first attempt gave a pattern similar to a combination of Figure A1b,c.

The $|gradB_z|$ distributions from simulations of all four configurations with separations $d = 2.92$ cm and 1.46 cm represent separations somewhat more than the 1.0 cm shielding distance described above. The values of *B1* were computed over the area shown in Figure A1e; the value of *B2* was extracted from the position 1 cm to the right of the position of each *B1* data point. The $|gradB_z|$ distributions were calculated from Equation (5). The results are shown in Figure A2.

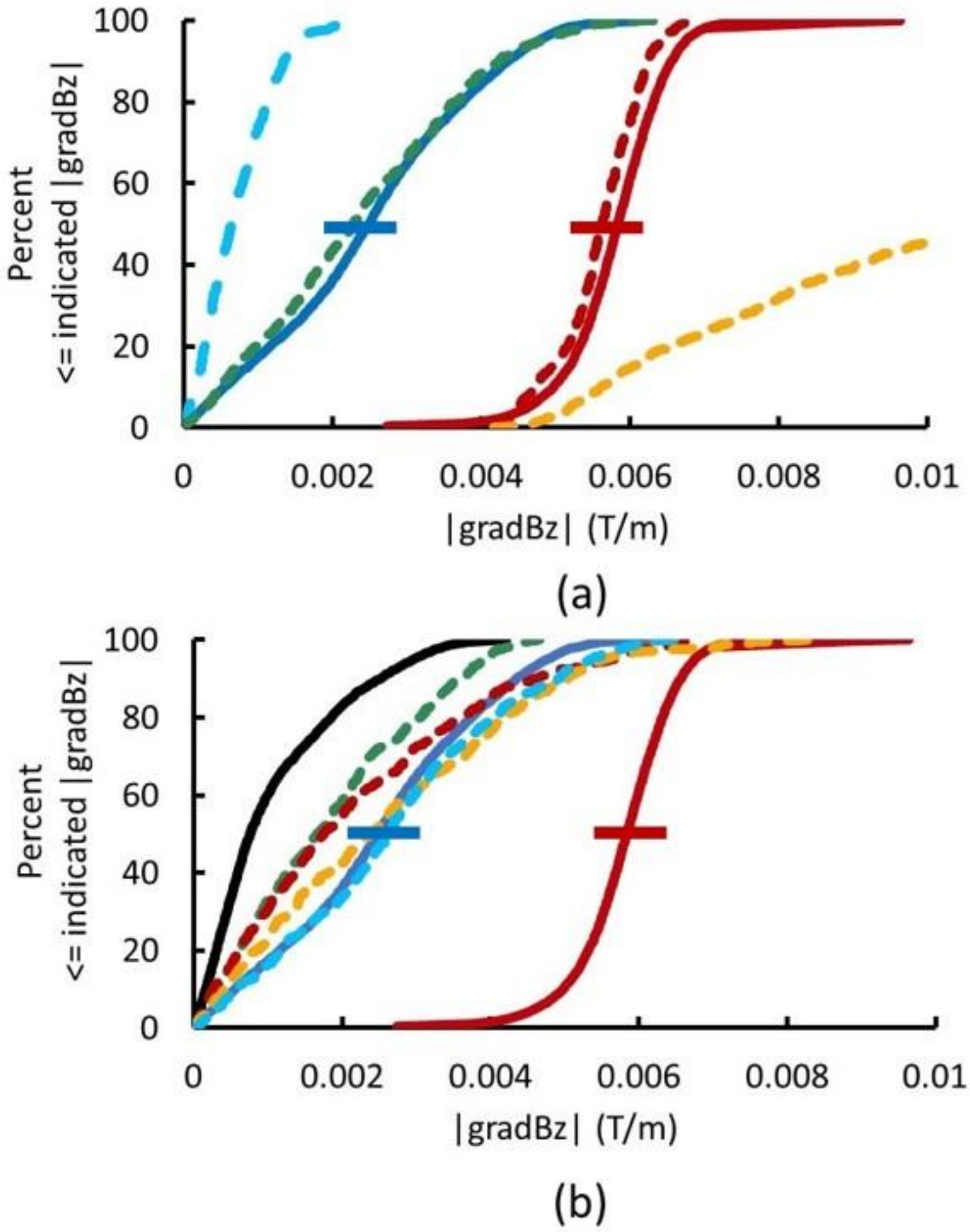

**Figure A2.** Representative |$gradB_z$| distributions from the 9-year old furnace 3 (solid blue), the first scan of a 41-year-old furnace 2 (solid red), control at 165 °C (black), and the |$gradB_z$| distributions from COMSOL simulations of the four patterns of magnetic moments shown in Figure A1: adjacent rows have opposite magnetic moments (light-blue dashes), clusters of three MQNs in a cell (green dashes), emulation in non-magnetic lattice (orange dashes), and all magnetic moments are in the same (+$x$) direction (red dashes). (**a**,**b**), respectively, show the results with $d$ = 1.46 cm and $d$ = 2.92 cm. The significance the error bars is described in the caption to Figure 4.

Figure A2a shows that only the pattern from $d$ =1.46 cm with all magnetic moments in +$x$ direction, i.e., Figure A1d, (red dashes) have nearly all points with |$gradB_z$| greater than some minimum value and consistent with the data from furnace 2. The emulation curve (orange dashes) intercepts the x-axis in the correct place but the slope is inconsistent with the furnace 2 data. Changing the MQN mass from 0.32 kg to 0.1 kg gives the correct slope but moves the intercept to |$gradB_z$| = 0.0012, which is inconsistent with the furnace-1 and furnace-2 data. Only the distribution from "all in +$x$ direction" has a shape consistent with the data from the furnace-scans shown in Figure 4a and illustrated by the data from furnace 2 (solid red line) in Figure A2a. The results of these simulations and of the emulation with neodymium magnets in a steel plate, as discussed above, support the "all in the +$x$ direction" as the best approximation to the pattern of magnetic moments inside the steel furnace bottom. Simulations of additional values of $d$ with the "all in +$x$ direction" pattern are shown in Figure A3.

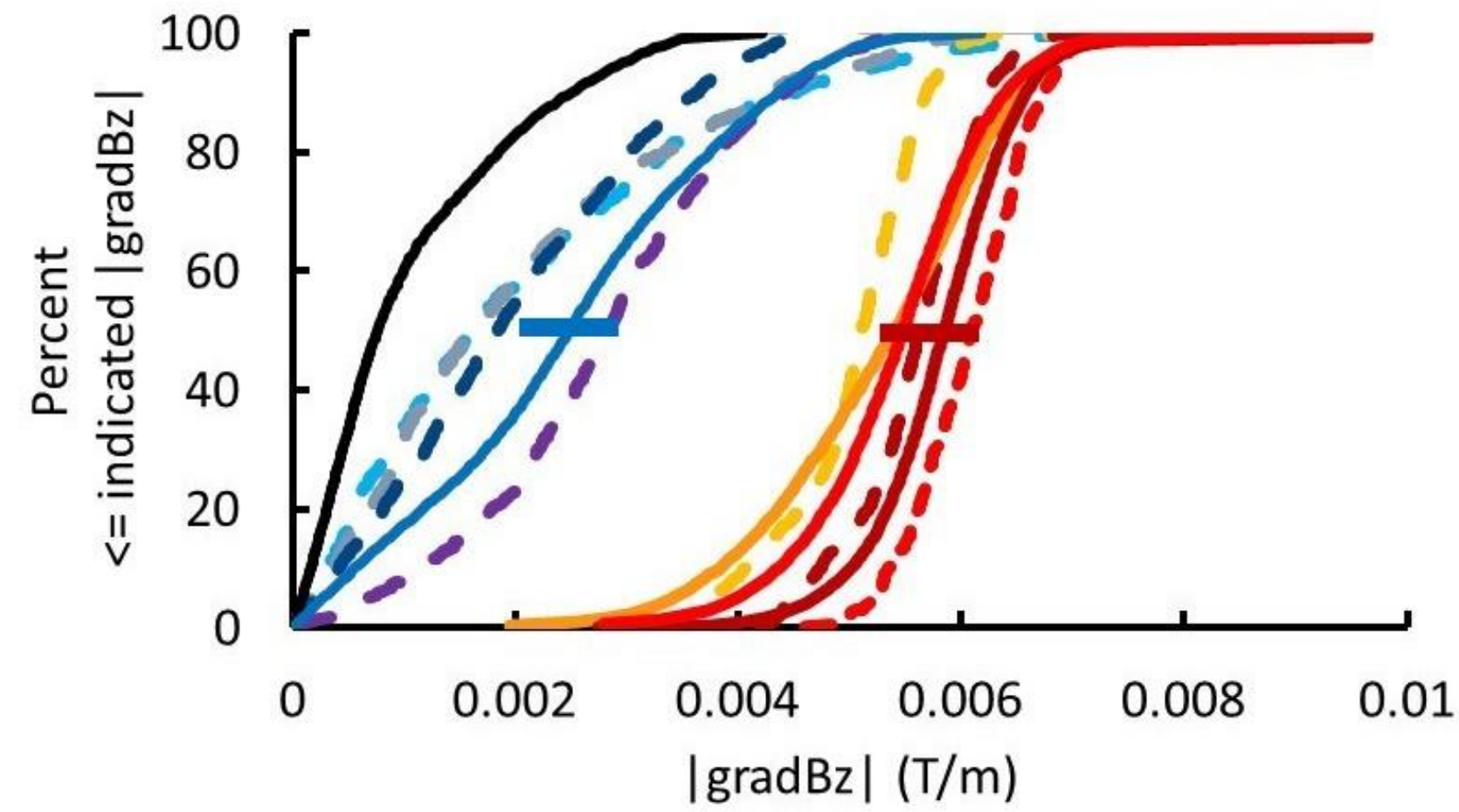

**Figure A3.** Comparison of data from furnace scans and COMSOL simulations of those scans as a function of MQN spacing *d* for MQN mass of 0.32 kg and with all MQNs' magnetic moments aligned in the +*x* direction. Solid lines show data: control at 165 °C (black), furnace 3 after 9 years of operation (blue), furnace 1 after 41 years of operation (orange), furnace 2 after 41 years of operation (dark red), repeat scan of furnace 2 showing variation in measurement technique (bright red). Dashed lines show simulation results as a function of MQN spacing *d*: with *d* = 3.0 cm (light blue), *d* = 2.5 cm (gray), *d* = 1.95 cm (dark blue), *d* = 1.9 cm (purple), *d* = 1.5 cm (orange), *d* = 1.46 cm (dark red), and *d* = 1.43 cm (bright red with small dashes). The significance the error bars is described in the caption to Figure 4.